\documentclass[letterpaper,notoc]{JHEP3}
\usepackage{amsmath, amsfonts}
\usepackage{yfonts}
\usepackage{array}
\usepackage{cite}
\usepackage{graphicx}

\newcommand{\la}{\langle}
\newcommand{\ra}{\rangle}
\newcommand{\Tr}{{\rm Tr}}
\newcommand{\be}{\begin{equation}}
\newcommand{\ee}{\end{equation}}
\newcommand{\bea}{\begin{eqnarray}}
\newcommand{\eea}{\end{eqnarray}}
\newcommand{\nl}{\nonumber \\}

\newcommand{\sgn}{{\rm sgn}\,}
\newcommand{\bp}{{\bf p}}
\newcommand{\bq}{{\bf q}}
\newcommand{\br}{{\bf r}}

\newcommand{\bv}{{\bf v}}
\newcommand{\dv}{\frac{d\Omega_v}{4\pi}}

\newcommand{\OO}{{\cal O}}

\newcommand{\Ref}[1]{{\rm Eq}.~(\ref{#1})}

\def\Im{\operatorname{Im}}
\def\Re{\operatorname{Re}}
\newcommand{\Nfour}{${\cal N}{=}4$}

\def\gsim{\mbox{~{\raisebox{0.4ex}{$>$}}\hspace{-1.1em}
	{\raisebox{-0.6ex}{$\sim$}}~}}
\def\lsim{\mbox{~{\raisebox{0.4ex}{$<$}}\hspace{-1.1em}
	{\raisebox{-0.6ex}{$\sim$}}~}}
\def\sss{\scriptscriptstyle}

\def\ch{C_{\!\sss H}}

\def\da{d_{\sss A}}
\def\ta{T_{\!\sss A\,}}

\def\dh{d_{\sss H}}
\def\nf{\mbox{$N_{\rm f}$}}
\def\nc{\mbox{$N_{\rm c}$}}
\def\mD{m_{\!\sss D}}
\def\mS{m_{\!\sss S}}
\def\gs{g_{\rm s}}
\def\alphas{\alpha_{\rm s}}
\def\Eq#1{{\rm Eq}.~(\ref{#1})}
\def\q{{\bf q}}
\def\r{{\bf r}}

\def\p{{\bf p}}
\def\putbox#1#2{{\epsfxsize=#1\textwidth \epsfbox{#2}}}
\def\putbox#1#2{\includegraphics[width=#1\textwidth]{#2}}
\def\centerbox#1#2{\centerline{\includegraphics[width=#1\textwidth]{#2}}}

\def\SYM{({\sss SY\!M\!})}
\def\QCD{({\sss QCD})}
\def\dksym{\delta\kappa^{\SYM}}
\def\dcsym{\delta C^{\SYM}}

\def\D{\displaystyle}

\title{Heavy quark diffusion in QCD and \Nfour\ SYM
at next-to-leading order}

\author{Simon~Caron-Huot and Guy~D.~Moore \\
    McGill University Dept.\ of Physics, 3600 rue University,
    Montr\'eal QC H3A 2T8 Canada \\ 
    E-mail: \email{scaronhuot@physics.mcgill.ca},
    \email{guymoore@physics.mcgill.ca}
}

\date{\today}

\abstract{
We present the full details of a calculation at next-to-leading order of
the momentum diffusion coefficient
of a heavy quark in a hot, weakly coupled, QCD plasma.
Corrections arise at $\OO(\gs)$; physically they represent
interference between overlapping scatterings,
as well as soft, electric scale ($p\sim gT$) gauge field
physics, which we treat using the hard thermal loop
(HTL) effective theory.
In 3-color, 3-flavor QCD, the momentum diffusion constant of a
fundamental representation heavy quark at NLO is
$\kappa = \frac{16\pi}{3} \alphas^2 T^3 ( \ln \frac{1}{\gs} +
0.07428 + 1.9026 \gs)$.
We extend the computation to a heavy fundamental representation
``probe'' quark in large $\nc$, \Nfour\ Super Yang-Mills theory, where the
result is
$\kappa^{\SYM} = \frac{\lambda^2 T^3}{6\pi} \left(
\ln \frac{1}{\sqrt\lambda} + 0.4304 + 0.8010 \sqrt{\lambda} \right)$ (where
$\lambda=g^2 \nc$ is the t'Hooft coupling).
In the absence of some resummation technique, the convergence of
perturbation theory is poor.
}

\preprint{}


\begin{document}

\section{Introduction}

In the earliest stages of the Big Bang the universe was a relativistic
plasma.  We can now produce such a relativistic plasma in the lab via
heavy ion collisions.  In both cases the plasma is transient--in the
early universe it lasted only around $10^{-6}$ seconds and in a heavy
ion collision it lasts little more than $10^{-23}$ seconds.  Since a
system which remains always in equilibrium leaves essentially no traces
of its earlier state, the most interesting physics in both situations is
nonequilibrium physics.  And in the early universe at temperatures above
10's of GeV [relevant for electroweak baryogenesis
\cite{baryo}, leptogenesis \cite{lepto}, gravitino production
\cite{gravitino_prod}, moduli production and destruction,
and other relics] that plasma was weakly
coupled.  The same is true in principle for the early development of
extremely high energy heavy ion collisions, though it is an open
question whether this is a reasonable treatment at available energies.

The natural language to study such plasmas is nonequilibrium quantum
field theory.  For plasmas near equilibrium (relevant for most of these
interesting problems) one can study linear deviations from (local)
equilibrium by studying unequal-time equilibrium correlations and using
linear response theory \cite{Kubo}.
The tools for defining equilibrium finite temperature field theory and
for performing weak coupling expansions have been known for over 40
years \cite{Matsubara,Keldysh}.  Nevertheless, our ability to calculate
thermal and nonequilibrium phenomena is surprisingly immature.
For thermodynamical
properties we now understand how to compute
perturbatively very well; for instance, the best known quantity,
the thermodynamic pressure, is known past fifth order
\cite{ArnoldZhai,BraatenNieto,gsixth}.  We also understand that to
compute real-time processes we need to perform a resummation of certain
plasma effects via the so-called Hard Thermal Loops (HTL's)
\cite{BraatenPisarski}.  

However surprisingly few gauge invariant
real-time correlation functions have been computed even at leading
order, and even fewer are known beyond this level.  In particular,
transport coefficients -- shear viscosity, baryon number diffusion,
electrical conductivity, heavy quark diffusion, bulk viscosity, and so
on -- are of considerable importance, since they describe the relaxation
of a system which is relatively close to equilibrium.
However, even leading order
calculations of these quantities only became available quite recently,
and {\em none} of them are known beyond leading order.

This is a major gap in our understanding of finite temperature field
theory.  In particular, it has been known for some time that the rate of
convergence of the perturbative series in the thermal theory can be
much worse than in the vacuum theory.
For instance, Braaten and Nieto \cite{BraatenNieto} argued that
the convergence of perturbation theory was poor unless $\alphas\lsim 0.1$,
due to physics at the so-called electric screening scale
$\sim gT$ (where the loop expansion really is only an expansion
into powers of $g$.)  It may be possible to rescue, or at least improve,
the convergence of perturbation theory using various resummation
techniques \cite{BraatenStrickland,BlaizotIancuRebhan}.  However neither
of these issues has been explored for dynamical (unequal time)
quantities.  In this context exactly one interesting gauge invariant
quantity is known
at next-to-leading order; the deeply virtual dilepton production rate
\cite{Majumder}.  However this quantity involves short physical time
scales and is not sensitive to the electric $gT$ scale, and is therefore
not representative of the other transport coefficients, which are.  For
these ``soft-sensitive'' quantities,
we do not know what a next-to-leading order computation
involves, we do not know how well the series will converge, and we do
not know what resummation techniques might be available or how well they
may work.

The simplest such quantity is the heavy quark diffusion coefficient,
first computed at leading order in \cite{braatenthoma}.
This is a quantity of phenomenological interest, which
sets the rate at which the velocity of a heavy quark equilibrates
with that of its environment.
In a recent Letter \cite{CHM1}
we have presented the result of a computation of the heavy quark
diffusion coefficient in full QCD to next-to-leading order.  Here we
present the full details of this calculation, as well as its extension
to \Nfour\ super-Yang-Mills theory (\Nfour\ SYM).  An
outline of the paper is as follows.  In the next section we summarize
the problem, discuss the relevant physics, and present the results.
Then Section \ref{sec:effective} presents our approach (HTL effective
field theory) and reviews the leading order calculation.
Section \ref{sec:calc} presents the body of the calculation
of the NLO diffusion coefficient, and Section \ref{sec:SYM} extends this
treatment to \Nfour\ Super-Yang-Mills theory.  A technical
appendix discusses the analytical properties of some of the functions we
encounter in the calculation.

\section{Overview and results}
\label{sec:overview}

\subsection{Definition of heavy quark diffusion}

A heavy quark, $M \gg T$, in or near equilibrium has a typical momentum
squared $\p^2 \sim MT \gg T^2$
large compared to the plasma scale and it therefore takes a
parametrically long time for the momentum to change appreciably.  This
means that momentum changes accumulate from many uncorrelated ``kicks,''
so on long time scales $p$ will evolve via Langevin dynamics,
\be 
\frac{d p_i}{dt} = -\eta_D\, p_i + \xi_i(t) \,,\quad 
\la \xi_i(t) \xi_j(t')\ra = \kappa\,\delta_{ij} \delta(t-t') \,.
\label{Langevin}
\ee
The relaxation rate $\eta_D$ and the momentum diffusion coefficient
$\kappa$ are
related by a fluctuation-dissipation relation,
$\eta_D = \frac{\kappa}{2MT}$,
which follows on general thermodynamical grounds.
Thus the dynamics of the nonrelativistic heavy quark
is completely set by the single parameter $\kappa$, which we
compute to next-to-leading order.

\subsection{Qualitative origin of NLO effects}
\label{sec:physics}

\FIGURE{
\centerbox{0.5}{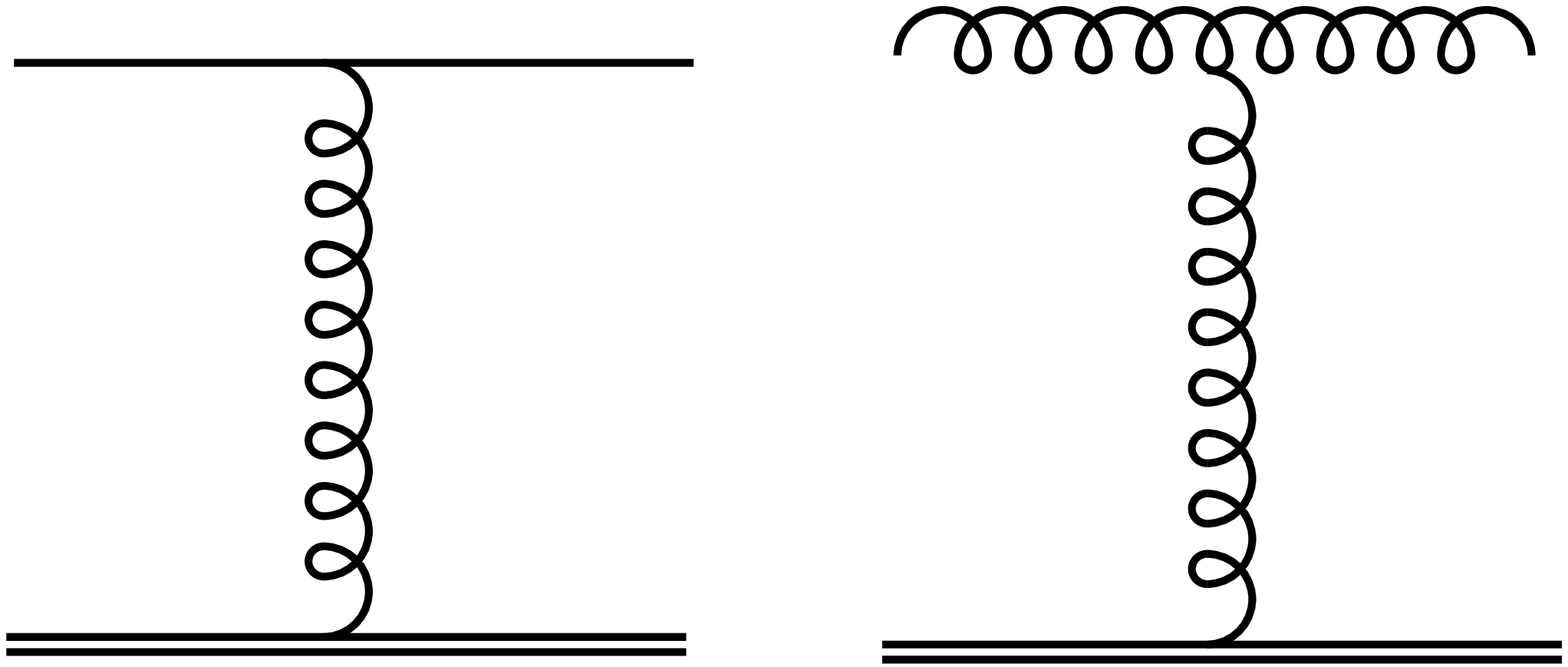}
\caption[Leading order scattering]
{\label{fig:LOscattering} Processes responsible for heavy quark
  diffusion at leading order:  Coulombic scattering between a heavy
  quark (double line) and either a quark or a gluon.}
}

At leading order in the weak coupling expansion,
the momentum diffusion coefficient is set by the t-channel
Coulomb scattering processes illustrated in Fig.~\ref{fig:LOscattering},
in which the scattering target can be a light quark or a gluon
(Compton-like processes are suppressed for nonrelativistic heavy
quarks.)  An important feature of weakly coupled plasmas, relativistic
and nonrelativistic, 
is that the total rate for Coulomb scattering is quadratically divergent
in the limit of small momentum transfer $|\bq|$
\cite{lifschitz}.
Such a divergence is of course unphysical in a medium of charged
particles, and is cut off by screening effects%
\footnote{
    In relativistic plasmas there still remains a logarithmic
    divergence due to un-screened magnetic scatterings, but this cancels
    in physical quantities and will not concern us here.
    }
at the momentum scale $q\sim \mD\sim gT$.

The momentum diffusion coefficient $\kappa$ is sensitive
not to the total scattering rate, but to the weighted average,
\be \kappa \equiv \frac{1}{3} \int d^3q \frac{d\Gamma(q)}{d^3q} q^2\,,
\label{schematickappa}
\ee
where $d\Gamma(q)/d^3q$ denotes the differential probability per unit
time for the momentum of the heavy quark to change by $\bq$.
The two additional powers of $q^2$ present in \Eq{schematickappa}
reduce the quadratic divergence of the total rate $\int d^3q d\Gamma/d^3q$
to a logarithmic divergence, cut off at $q\sim \mD$.
The logarithm reflects the fact that the heavy quark momentum diffuses
due to a range of momentum transfers, from the
many soft scattering events, which individually
have $q\sim gT$ but occur on a rate $\Gamma_{\rm soft}\sim g^2T$, to the
rare hard scattering events with $q\sim T$, occurring on a rate
$\Gamma_{\rm hard}\sim g^4T$.

Usually interaction corrections involve powers of $g^2$, and for large
momentum transfer processes this is true.  But the presence of the
Coulombic divergence, cut off by screening effects, means that
the details of soft momentum exchange and plasma screening are relevant
to heavy quark diffusion already at leading order.  Now consider the
contribution from gluons to the Debye screening scale
\cite{weldonold}:
\be 
\mD^2 = 4 \ta g^2 \int \frac{d^3k}{(2\pi)^3} \frac{n_B(k)}{k} +
\mbox{(fermion contribution)}\,.
\label{schematicmd}
\ee
Here $\ta = \nc = 3$ is the trace normalization of the adjoint
representation and $n_B$ is the Bose distribution function.
At small $k$ the integral behaves like $\sim g^2T\int d|k|$, due to the
singular nature of the Bose-Einstein function $n_B(k)\sim T/k$.
Thus the contribution from gluons with $k \sim gT$ represents $\OO(g)$ of the
total strength of plasma screening.  

However, the soft gluon contribution is not
computed correctly by the above expression.  The derivation of
\Eq{schematicmd} assumed free massless propagation of the particles
responsible for screening, but gluons with $\OO(gT)$ momenta
themselves experience $\OO(1)$ screening effects.
Therefore one must recompute that part of plasma screening
which arises from the $\OO(gT)$ gluons.  This calculation is complicated
by the fact that interactions between soft gluons are also strongly
modified by plasma effects, described by the HTL
effective theory \cite{BraatenPisarski,FrenkelTaylor}.
Thus, since a relative $\OO(g)$ fraction of \Eq{schematicmd} arises
from soft gluons, a correct treatment of the effect they produce
will produce an $\OO(g)$ correction to the physics which cuts off
the infrared logarithm in $\kappa$, and therefore an $\OO(g)$ correction
to the result.

For similar reasons, in an $\OO(g)$ fraction of soft scattering events,
the plasma particle which strikes the heavy quark is itself a soft gluon
with momentum $\sim gT$.  But the dispersion and spectral weight of such
gluons are strongly modified by the plasma and this contribution to the
``target density'' must also be reconsidered.

\FIGURE{
\centerbox{0.7}{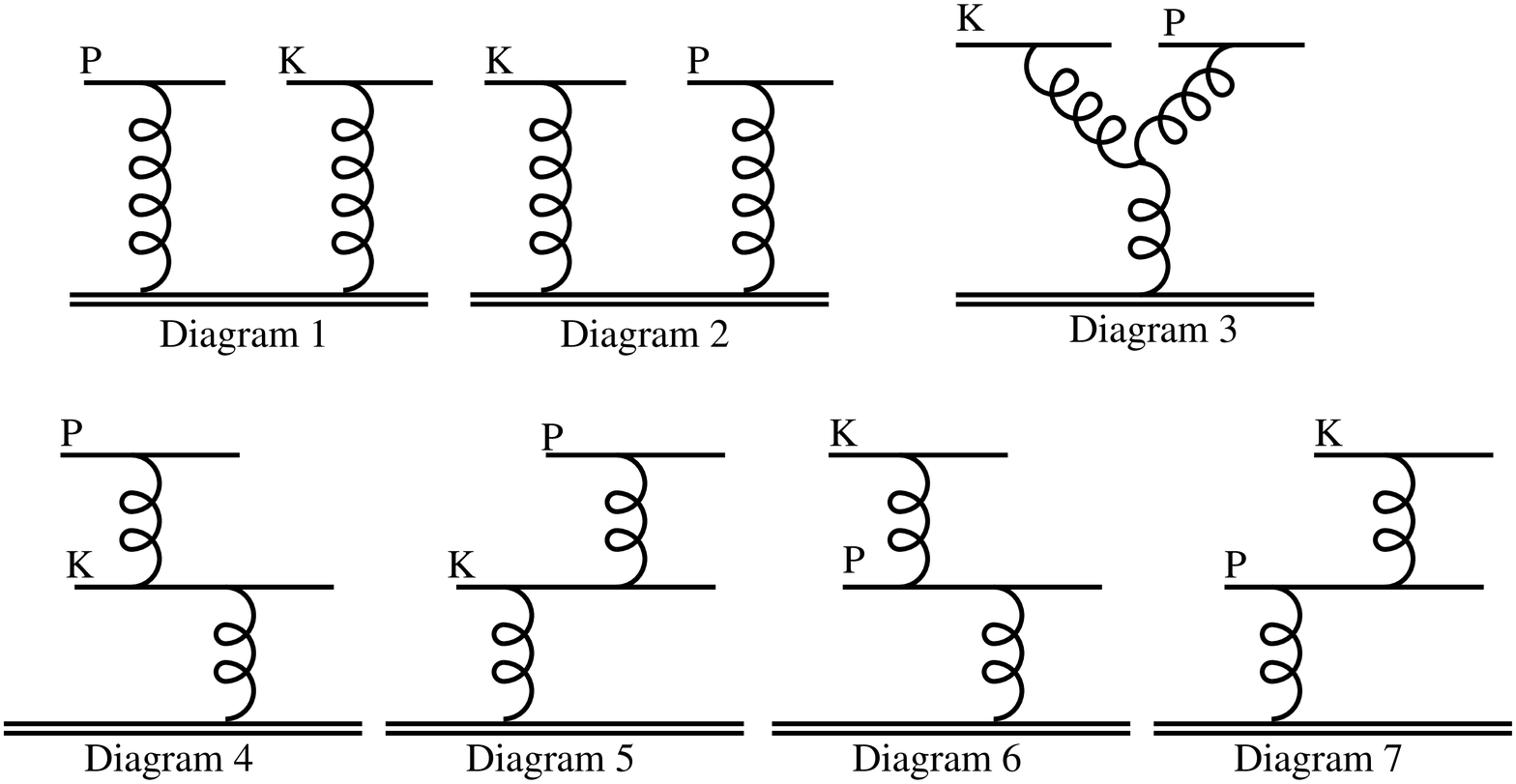}
\caption[Overlapping scattering events]
{\label{fig:overlap}
Diagrams which can interfere when two scattering processes overlap.
Here $K$ and $P$ are the initial momenta of two distinct hard particles,
which can be quarks or gluons.}
}

Other NLO corrections can be expected to arise from
interference between overlapping scattering events,
as illustrated in Fig.~\ref{fig:overlap}.
Since the \emph{total} scattering rate is $\sim g^2T$,
both for the heavy quark and for the light particles which scatter it,
and since small angle scatterings can have a duration
up to $\sim 1/gT$ (this can be read off, for instance, from the virtuality
of the exchanged gluon), an $\OO(g)$
fraction of scattering events overlap with another scattering.
Thus the individual diagrams in Fig.~\ref{fig:overlap} are only
down by $\OO(g)$ relative to the leading processes.  Corresponding
virtual corrections to the leading order scattering process will
naturally also arise at this order.
However, the appropriate question to ask
is not about the probability for scattering events to overlap,
but rather how much do they interfere with each other.
In QED, the small angle scattering of a particle occurring
during another scattering event
has little impact on that event, and correspondingly
one finds a parametric cancellation between diagrams 1 and 2
of Fig.~\ref{fig:overlap} (and
between diagrams 4 and 5 and between diagrams 6 and 7, and among the
associated virtual processes; diagram 3 is absent altogether in QED.)
However, in QCD, instead of a cancellation one gets a commutator
of group theory factors: the very frequent soft scatterings which
occur in the plasma \emph{do} matter, because they change the colors
of the particles.

Note however that this source of $\OO(g)$ NLO corrections and the
preceding one are not clearly distinct.  Indeed, diagram 3 of
Fig.~\ref{fig:overlap} can be understood as the special case of
the diagram in Fig.~\ref{fig:LOscattering}, where the external gluons
are soft and in the Landau cut.  This suggests that, to make the
qualitative discussion here more precise, we will need to perform a
careful diagrammatic approach based on power counting.  There is one
common feature of the sources for correction we have listed, however;
all involve the influence of soft gluons.
This observation suggests that the calculation may be rephrased in terms
of an effective theory of $gT$ scale physics, in which
the hard scale $\sim T$ has been integrated out.  This is precisely
Braaten and Pisarski's HTL effective theory
\cite{BraatenPisarski}.  Carrying out a careful
diagrammatic calculation within this effective theory is the
subject of the body of this paper; in the remainder of this section we
will present the results.

\subsection{Results:  QCD}

The squared matrix elements for the processes of
Fig.~\ref{fig:LOscattering}, summed over the initial and final
states of the light scattering targets and final states of the heavy
quark, and averaged over the initial states of the heavy quark,
have been evaluated in \cite{MooreTeaney}, yielding
\be \kappa^{\sss LO} \equiv \frac{g^4\ch}{12\pi^3}
\int_0^\infty k^2dk \int_0^{2k} \frac{q^3dq}{(q^2+\mD^2)^2}
\times \left\{\begin{array}{l} \D
N_c\,n_B(k)(1{+}n_B(k))      \left(2-\frac{q^2}{k^2} + \frac{q^4}{4k^2}\right) \\ \D
+N_f\,n_F(k)(1{-}n_F(k))\left( 2-\frac{q^2}{2k^2} \right)   \,,
\end{array} \right. \label{kappalo}
\ee
where $\ch=\frac43$ in QCD is the quadratic Casimir of the heavy quark
representation, and $\mD=\sqrt{1.5}gT$ in QCD with \nf=3 flavors of light quarks.
Formally taking $\mD \ll T$, the integral is dominated by $k \sim T$
and $q$ in the logarithmic range $\mD \lsim q \lsim T$.
The leading behavior in $g$ of \Eq{kappalo} can be obtained from
the leading behavior in $m_D^2/k^2$ of the $q$ integral.
Making room for the next-to-leading order correction $C$,
the result can be written:
\be
\kappa \!=\! \frac{\ch g^4 T^3}{18\pi} \!
      \left( \!\left[\nc{+}\frac{\nf}{2}\right] \!\!
\left[\ln\frac{2T}{\mD}{+}\xi\right] {+} \frac{\nf \ln 2}{2}
{+} \frac{\nc \mD}{T} C + \OO(g^2) \!\right) \,.
\label{defkappa}
\ee
Here $\xi = \frac{1}{2}-\gamma_{\sss E} + \frac{\zeta'(2)}{\zeta(2)}
\simeq -0.64718$. The leading order part of \Eq{defkappa}
was given explicitly in \cite{MooreTeaney} (it could also have been extracted
from the nonrelativistic limit of earlier results
\cite{braatenthoma,svetitsky}.)
The dependence of the next-to-leading order correction on physical
parameters is contained in the coefficient multiplying $C$, which itself
is a pure number:
all of the above-mentioned next-to-leading order corrections depend
on physical parameters in the same way as
an $\OO(\mD/T)$ fraction of the gluon contribution to $\kappa^{\sss LO}$.

Expression \Ref{kappalo} itself contains $\OO(g)$ corrections,
giving rise to a rather trivial contribution\footnote{
In \cite{CHM1} this contribution
was named $C_{\mbox{Eq.\,(4)}}$.} to $C$,
$C_{2\to2}=\frac{21}{8\pi}\simeq 0.8356$.
It arises wholly from the $k\sim gT$ region of the gluon contribution to
\Ref{kappalo},
where the result of the $q$ integration
is poorly described by the leading term of its $\mD^2/k^2$ expansion,
which was used to obtain the leading order behavior \Ref{defkappa}.
Although slightly tedious, the evaluation of $C_{2\to2}$ is
entirely straightforward and we do not present it here.
In section \ref{sec:calc} we compute the difference between
the full next-to-leading order momentum diffusion coefficient,
and what is already incorporated in $\kappa^{\sss LO}$, and obtain
$\tilde{C}\simeq 1.4946$.
Thus $C\equiv C_{2\to 2}+\tilde{C}\simeq 2.3302$.

\FIGURE{
\centerbox{0.7}{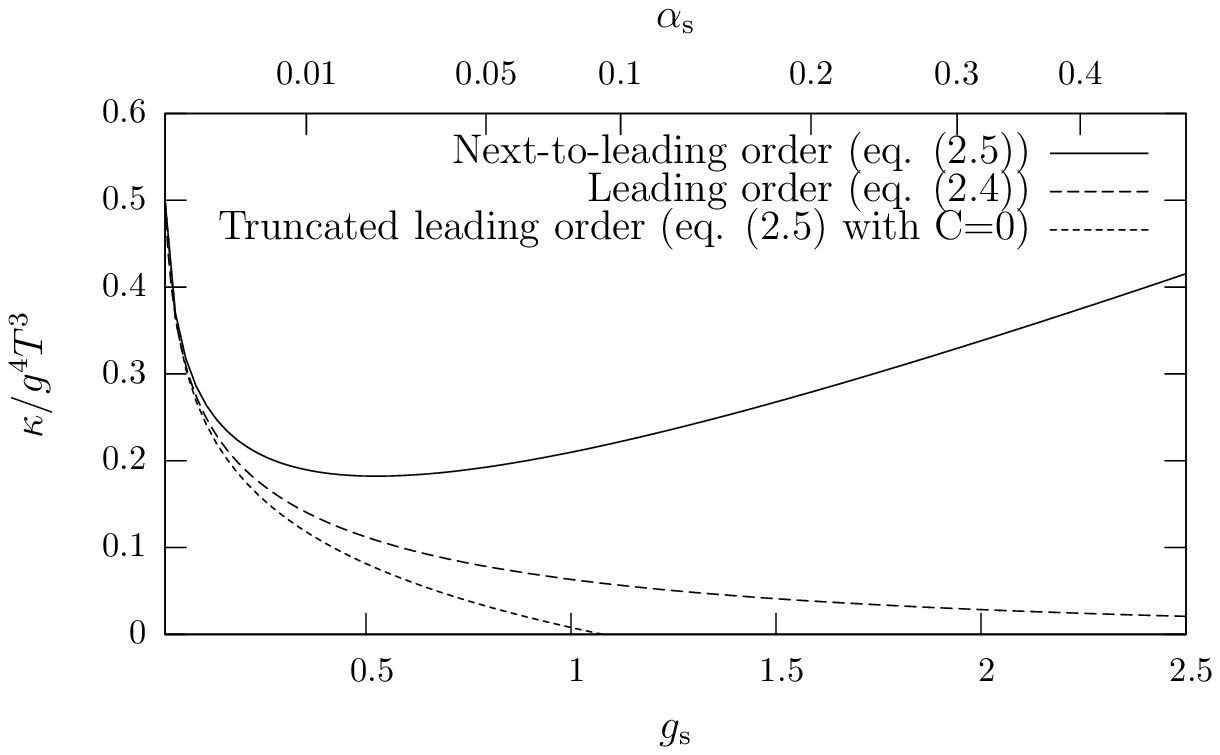}
\caption[Results in QCD]
{\label{fig:qcd}
Comparison of leading and NLO results for $\nf=3$ QCD
  as a function of coupling.}
}

Our result \Ref{defkappa} is plotted in Fig.~\ref{fig:qcd}.
A simple-minded estimate of the regime of validity of perturbation
theory can be given by equating the size of the correction to the size
of the leading-order result. What is usually referred to in the literature
as being the leading order result is \Ref{kappalo}, numerically integrated at a given
value of the coupling (this is the curve called ``leading
order'' in Fig.~\ref{fig:qcd}):
the correction becomes as large as this leading order result when
$\alphas\gsim 0.04$.
This suggests that at that point perturbation theory starts to get into trouble.
For this reason, and as should be clearly suggested by the plot, we do not
believe that our calculation can be directly used as an ``improvement''
to the determination of $\kappa$ in the context of heavy ion collisions, where
phenomenologically realistic values of the coupling are in the range
$\alphas\sim 0.3-0.5$.  Rather our results signal difficulties with the
approach.

Nevertheless we would not like to sound overly pessimistic and conclude
that our results signal
that no prediction beyond $\alphas=0.05$ is possible.
Rather, the real question now is how large the higher order corrections are,
and more pertinently, which parts of $C$ may
duplicate themselves in higher-order terms,
in some more or less predictable (and therefore resummable) fashion.

Consider for instance the difference between the two lowest curves of
Fig.~\ref{fig:qcd}, which is attributable to $C_{2\to2}$,
up to terms which are of yet higher order in the $\mD/T$
expansion of \Ref{kappalo}.
This contribution, which can be evaluated knowing only
the tree-level matrix elements with massless external states
(and HTL corrections resummed on the exchanged gluon),
is better described as an ``ambiguity'' in the
leading-order result rather than as a correction to it.
This ambiguity is large because the Coulomb scattering processes
against soft gluons (which give the small $k$ contribution to \Eq{kappalo})
are poorly described by the leading term of an $\mD/T$ expansion.
This is unrelated to the question
of whether these processes are correctly described by the right-hand
side of $\kappa^{\sss LO}$, which is the most pertinent question
to ask if we are concerned with higher order terms.
Actually, our calculation tells us that the effect of the
HTL changes in the dispersion relation and interaction strength of the soft
gluons is relatively modest,
essentially given by the pole-pole contribution
of section \ref{sec:finala}, of order $C_{(A),\rm pole-pole}\sim -0.20$.
Thus it appears that this region of phase space is not so poorly
described by \Ref{kappalo}, and that simply
defining this full expression
to be the leading order result
should provide a reasonable resummation
of the contribution $C_{2\to 2}$.

Along the same lines, there is another contribution to $C$ which has a
simple physical interpretation and which would be easy to include into the
leading order calculation.  As we will discuss in Subsection
\ref{sec:re}, just over half of the remaining correction arises from a
shift in the (real) Debye screened propagator $1/(q^2+\mD^2)$ appearing
in \Eq{kappalo}.  This can be understood as an NLO momentum dependence
in the Debye screening mass $\mD^2$.  This momentum dependence can be
resummed in a few ways.  One way would be to solve for it in the 3D
Euclidean effective theory nonperturbatively or via a gap equation
(though this approach appears to be special to heavy quark diffusion,
where the exchange momentum is strictly spatial).  Another method would
be to replace $\mD^2$ in \Eq{kappalo} with the full momentum-dependent
leading-order self-energy at general $q$ (though this procedure does not
appear to be gauge invariant).  

For these reasons we think with some optimism
that two thirds of the difference between the lowest and highest curves
in Fig.~\ref{fig:qcd} can be absorbed in a relatively simple systematic
resummation scheme, and at most only one third
represents complicated physics that will be really difficult to resum.
Such a resummation scheme might
then extend the reach of perturbation theory to $\alphas \sim 0.15$ or
so--high enough for almost all cosmological applications, though still
not enough to be much use for heavy ion physics.  Clearly such issues of
resummation are an interesting problem for future work.

\subsection{\Nfour\ super Yang-Mills}
\label{sec:resSYM}
Results for heavy quark diffusion in the \emph{strong}
coupling regime of \Nfour\ super Yang-Mills (SYM) have been obtained
in the literature,
exploiting the AdS/CFT correspondence \cite{CKKKY,CasalderreySolana}.
It seems interesting to study the heavy quark momentum diffusion
coefficient also at weak coupling in this theory, for the purpose
of comparison between the two regimes and for comparison between this
theory and ordinary QCD.
The leading order result at weak coupling
has previously been given \cite{vuorinenchesler}, and here
we present the next-to-leading order correction it receives.%
\footnote{
    We present our results at leading order in the large $\nc$
    expansion.  Strictly speaking, the theory at finite
    $g^2=\lambda/\nc$ and with
    added fundamental matter has a Landau pole; however it is valid
    perturbatively.  To generalize \Eq{SYMLO} and \Eq{kappaSYM} to
    finite $\nc$, multiply the righthand sides by $2\ch/\nc$.
    }

We begin with a brief description of \Nfour\ SYM and of heavy quarks
in this theory.
In addition to the gauge field $A_\mu$,
\Nfour\ SYM contains four Weyl fermions and six real scalars, all
transforming under the adjoint representation of the gauge group.
The theory contains a single dimensionless coupling constant $g$,
which sets the strength of the gauge, Yukawa and scalar interactions;
the Lagrangian is completely determined by the supersymmetry \cite{refSYM}.
The strong coupling results are obtained in the large \nc\ limit
of the theory with gauge group SU(\nc), and for this reason
we will express our results in terms of the t'Hooft coupling
$\lambda\equiv g^2 \nc$.
What is meant by a ``heavy quark'' in this theory is
an ${\cal N}{=}2$ massive hypermultiplet added to it, transforming under
the fundamental representation of the gauge group.
In terms of the ${\cal N}=2$ field content of \Nfour\ SYM,
this heavy hypermultiplet is minimally coupled to the ${\cal N}=2$ gauge
multiplet of the theory, but is not directly coupled to the massless
matter hypermultiplet.
This is the conventional setup employed in AdS/CFT studies \cite{CKKKY}.

\FIGURE[t]{
\centerbox{0.5}{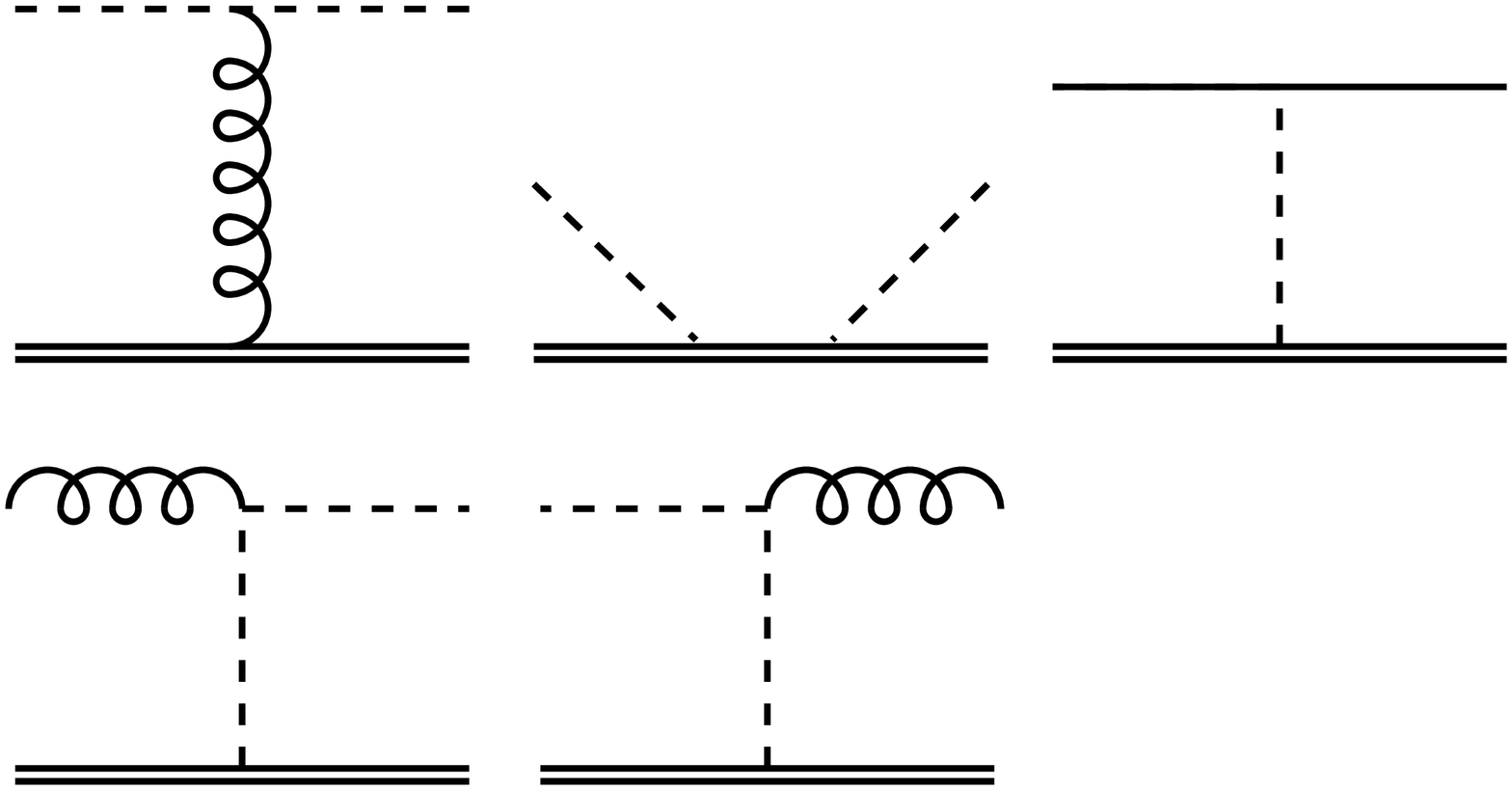}
\caption[New leading order SUSY graphs]
{\label{fig:LOSYM}
Leading order diagrams involving scalars, which are present in \Nfour\
SYM but not in QCD.}
}

In the large mass $M \gg T$ limit,
processes which would change the identity
of the heavy particles, from heavy quarks to heavy scalars and
vice-versa, are suppressed\cite{vuorinenchesler}. Thus the heavy
fermion carries an approximately conserved U(1) charge and it makes
sense to speak about its momentum diffusion coefficient,
with no reference to its scalar superpartners.
In addition to the scattering processes depicted in
Fig.\ \ref{fig:LOscattering}, in this theory at leading order there are scattering
processes involving light scalars, depicted in Fig.\
\ref{fig:LOSYM}.
Including these processes, the leading order
momentum diffusion coefficient $\kappa$
can be written \cite{vuorinenchesler}:
\bea \kappa^{\rm LO}_{\SYM} &=&
\frac{\lambda^2}{24\pi^3}\int_0^\infty k^2dk \int_0^{2k} q^3dq
\nl && \times 
\left\{\begin{array}{l} \D
n_B(k)(1+n_B(k)) \left(2-\frac{q^2}{k^2} +
\frac{q^4}{4k^4}\right)/(q^2+\mD^2)^2
\\ \D
+n_B(k)(1+n_B(k)) \left[ \frac{5}{(q^2+\mD^2)^2} + 
\left(\frac{1}{q^2+\mD^2}-\frac{1}{2k^2}\right)^2 \right]
\\ \D
+4n_F(k)(1-n_F(k)) \left(2-\frac{q^2}{2k^2} \right)/(q^2+m_D^2)^2
\\ \D
+2n_B(k)(1+n_B(k))
\left(\frac{q^2}{k^2}-\frac{q^4}{4k^4}\right)/(q^2+\mS^2)^2
\\ \D
+2n_F(k)(1-n_F(k)) \frac{q^2}{k^2} /(q^2+\mS^2)^2 \,,
\end{array} \right. \label{SYMLO}
\eea
where $m_D^2=2\lambda T$ and $m_S^2=\lambda T$.
The first line describes Coulomb scattering against gluons,
the second line describes Coulomb scattering against five real scalars
and Coulomb plus Yukawa-Compton scatterings against one real scalar%
\footnote{Apparently the Yukawa-Compton scattering processes against
  scalars (the second diagram in Fig.~\ref{fig:LOSYM})
  were not included in the calculation of Vuorinen and Chesler
  \cite{vuorinenchesler}. This caused an error in their determination of
  the constant term in \Eq{kappaSYM}; they found $\frac{7}{12}$ rather
  than $\frac{1}{2}$.}.
The third line is Coulomb scattering of fermions,
the fourth line is the scalar-mediated conversion
of light gluons to light scalars and vice-versa, and the last line contains
the scalar-mediated scatterings against light fermions.
The integrals of \Eq{SYMLO} were evaluated to leading order in
  $\mD/T\sim \sqrt{\lambda}$
in \cite{vuorinenchesler}; making room for the next-to-leading order
contribution the result can be written:
\be
\kappa^{\SYM} = \frac{\lambda^2T^3}{6\pi} \left(
\ln\frac{2T}{\mD}+\xi {+} \frac{1}{2}  +\frac13\ln 2 + \frac{\sqrt{2\lambda}}{6} C^{\SYM}
 + \OO(g^2) \!\right) \,,
\label{kappaSYM}
\ee
with $\xi$ as defined below \Eq{defkappa}.
As for QCD, a rather trivial contribution
$C_{2\to2}^{\SYM}=\frac{15}{2\pi} - \frac{3}{\pi\sqrt2}\simeq 1.7121$ to $C^{\SYM}$
arises from the expansion of \Eq{SYMLO} to next-to-leading order in
$\sqrt{\lambda}$, coming from the $k\sim \sqrt{\lambda}T$ region of processes involving
external bosons. Another part of $C$ is precisely the same as the ``difficult'' part
of the QCD calculation, $\tilde{C}^{\QCD}\simeq 1.4946$. The remainder
$\tilde{C}^{\SYM}\simeq 0.19172$
is calculated in section \ref{sec:SYM}. Thus
$C^{\SYM}=C_{2\to2}^{\SYM}+\tilde{C}^{\QCD} +\tilde{C}^{\SYM}\simeq 3.3984$.

\FIGURE[t]{
\centerbox{0.6}{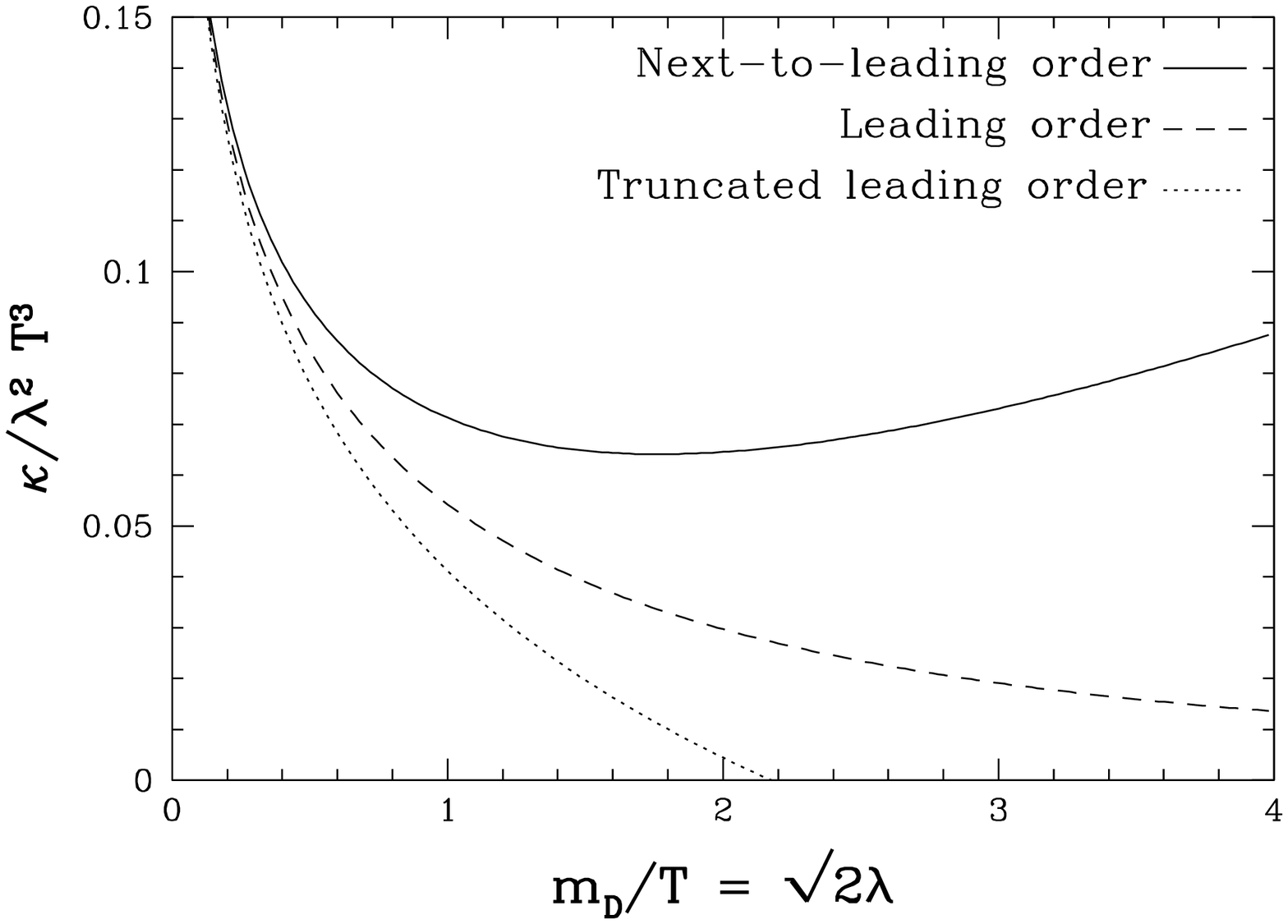}
\caption[SYM results]
{\label{fig:SYM}
Comparison of next-to-leading and leading order results for heavy quark
diffusion in \Nfour\ SYM theory.}
}

This result is plotted in Fig.\ \ref{fig:SYM}.  As the figure shows, the
convergence of the series is somewhat better than in QCD.
Viewed as a function of the t'Hooft coupling $\lambda$ the correction is
100\% of the leading value for $\lambda \simeq 1.71$, to be compared with
the QCD value of $\alphas=0.033$, which is $\lambda=1.26$.
Viewed as a function of the Debye screening scale, the comparative
convergence in SYM would seem even better; the 100\% correction point occurs for
$\mD = 1.85 T$, whereas it is $\mD = \sqrt{\frac{3}{2}} gT= 0.79 T$.
However, estimates of this kind can sometimes be deceptive:
the fact that at fixed $\mD/T$ the NLO correction in SYM
is comparatively smaller than in QCD can be mostly attributed to the leading order
result being stronger in SYM, due to its larger number of matter fields
(which also scatter with a larger group theory factor, being in the
adjoint representation.)
However, this additional physics that is present in SYM suffers from relatively
modest NLO corrections, the most severe corrections still being associated
with soft gluons.  Thus, although the leading order result has a wider
range of validity in SYM than in QCD,
it seems reasonable to expect the range of validity of the
NLO correction itself to be roughly similar in SYM and QCD, in terms
of $\mD/T$.

\section{Effective theory and leading-order analysis}

\label{sec:effective}

Our first step towards a rigorous analysis of the
momentum diffusion coefficient $\kappa$ is
a non-perturbative definition \cite{CasalderreySolana} in terms
of a force-force (electric field-electric field) correlator
with Wilson lines connecting the electric fields:
\be 
\kappa \equiv \frac{g^2}{3\dh}
  \!\!\int_{-\infty}^\infty \!\!\!\!\!\!dt\:
\Tr_H \la W(t;-\infty)^\dagger \,E_i^{a}(t)t^a_H 
\,W(t;0) \,E_i^{b}(0)t^b_H \: W(0;-\infty) \ra\,. 
\label{correl1}
\ee
The trace runs over the representation of the gauge group
of the heavy quark, and the Wilson lines $W$ act on
this representation.
Intuitively, \Ref{correl1} is exactly the force-force correlator of
\Eq{Langevin}, with the forces given by electric fields and the Wilson
line representing the gauge rotation of the heavy quark due to
propagation, which ensures
gauge invariance.  Because of operator ordering
issues, the Wilson lines shown are not equivalent to connecting the $E$
fields with an adjoint Wilson line.
The Wilson lines also incorporate the effect of the heavy charge on the
plasma (which is why they must go back to time $-\infty$).  In general
they introduce nontrivial representation dependence
into the heavy quark diffusion constant,
and in fact such Wilson lines are
even required in QED (diffusion of ions in a QED plasma depends on
the ionic charge $Z$ in a more complicated way than $Z^2$ only because
of these Wilson lines, which account for the reaction of the plasma to
the presence of the charge).  However we will see that to the order we
work here, they can be replaced by an adjoint Wilson line.

Our approach is to calculate this correlation function within HTL
effective field theory.  This is an infrared effective description valid
below a cutoff scale $\sim T$ which describes gauge fields and fermion
fields, resumming into the Lagrangian certain $O(T^2)$ plasma effects.
Perturbatively this introduces corrections to the propagators and
vertices which become $O(1)$ at the scale $\sim gT$, which is a natural
scale in HTL effective theory.  The effective theory requires matching
to the full thermal theory in the UV, requiring counterterms both in the
Lagrangian parameters and in correlation functions such as \Eq{correl1}.
Perturbation theory within the HTL effective theory is expected to
converge in powers of $g$, which intuitively can be though of as the
usual factor $g^2$ times a Bose statistical factor evaluated at the
scale $gT$, $n_{\sss B}(gT) \sim 1/g$.  The HTL calculation can also
encounter infrared divergences, arising from the unscreened
low-frequency magnetic gluons.  The appearance of such an IR divergence
signals the breakdown of perturbation theory and the need for
nonperturbative information about the ultrasoft magnetic sector.  We
expect such IR divergences at some finite order in perturbation theory,
but this proves to be beyond the NLO level we consider here.%
\footnote{%
    We believe that most transport coefficients are sensitive
    to nonperturbative magnetic physics at $\OO(g^2)$, which is the
    relative contribution of these magnetic fields to transverse
    momentum diffusion for a
    moving particle ($v\sim 1$) in the soft electromagnetic fields of
    the HTL effective theory.  Because the heavy quark considered here
    has $v \ll 1$, we believe magnetic physics arises at a higher
    order than $\OO(g^2)$.}%

Let us proceed with the leading order calculation of the correlator in
\Eq{correl1}.
At leading order one may replace the Wilson lines with identity
operators and use the noninteracting form of the electric field
correlator,%
\footnote{
    We work in $[{-}{+}{+}{+}]$ metric contentions and 4-vector
    potential $A^\mu = (A^0,A^i)$ with $A^0$ the usual scalar and $A^i$
    the usual vector potentials.
    }
$E_i = \partial_i A_0 - \partial_0 A_i$.  The time integral means we
need the result at zero frequency and so the $\partial_0 A_i$ piece does
not contribute%
\footnote{
    This does not apply in gauges such as temporal gauge where the zero
    frequency gauge boson propagator can display singularities.
    }.
Fourier transforming to the momentum basis, we rather immediately obtain
\be
\kappa_{\sss LO} = \frac{g^2 \ch}{3\da} \int \frac{d^3 q}{(2\pi)^3}
  q^2 \langle A_{0a}(\omega=0,q) A_{0a}(0) \rangle 
 = \frac{g^2 \ch}{3} \int \frac{d^3 q}{(2\pi)^3}
  q^2 G_{00}^>(0,q)  \,.
\ee
The Wightman propagator is to be evaluated within the HTL effective
theory.  Using the KMS condition we can express the Wightman propagator
in terms of the retarded propagator,
\be
G^>(\omega,q) = 2(n_{\sss B}(\omega){+}1) \Re G_{\sss R}(\omega,q)
\ee
which is given in the HTL effective theory, in strict Coulomb gauge
(which we use throughout), by
\bea
G_R^{00}(P)&=& \frac{i}{p^2+\Pi_R^{00}(P)}, \nl
G_R^T(P)&=& \frac{-i}{P^2 + \Pi_R^T(P)}. \label{HTLprop}
\eea
with
\bea
\Pi_R^{00}(P) &=& m_D^2\left[1 - \frac{\eta}{2}
  \ln\left(\frac{|1+\eta|}{|1-\eta|}\right)
+ \frac{i\pi \eta}{2} \theta(1-\eta^2) \right] \nl
\Pi^T_R(P) &=& m_D^2 \left[ \frac{\eta^2}{2} + \frac{\eta(1-\eta^2)}{4}
\ln\left(\frac{|1+\eta|}{|1-\eta|} \right)
-\frac{i\pi\eta(1-\eta^2)}{4} \theta(1-\eta^2) \right],
\eea
where $\eta \equiv p^0/p$. 
Therefore the leading order momentum diffusion coefficient is
\be
\kappa^{\sss LO} =
\lim_{\omega \rightarrow 0}
\frac{g^2 \ch}{3} \int \frac{d^3 q}{(2\pi)^3} q^2 \frac{2T}{\omega}
\frac{\pi \omega \mD^2}{2q(q^2+\mD^2)^2}
= \frac{g^2 \ch \mD^2}{6\pi} \int \frac{q^3 dq}{(q^2+\mD^2)^2} \,.
\label{kappalo2}
\ee
This integral is UV log divergent, indicating the need to perform a
matching calculation.  This is done by finding the result in the full
theory, which gives \Eq{kappalo} (without the $\mD^2$ terms in the
denominator).  In the range $gT \ll q \ll T$ the two calculations agree;
performing the $k$ integration in \Eq{kappalo} treating $q \ll k$
reproduces \Eq{kappalo2}. 
The matching should be performed in some way
so that the UV region is equivalent to \Eq{kappalo} and the IR region is
equivalent to \Eq{kappalo2}.
One way of doing this would be to
compute both
\Eq{kappalo} (without $\mD^2$ factors in denominators) and
\Eq{kappalo2} each in dimensional regularization and add them; the
$1/\epsilon$ factors will cancel and the finite parts will give a
consistent leading order result \cite{Brown}.  Alternately one can
simply insert $\mD^2$ factors in the denominators in \Eq{kappalo} (as
has already been done) so that one expression is appropriate in the IR
and UV.  The error thus introduced at large $q$
is only NNLO ($\OO(g^2)$) and will not interfere with our NLO calculation.


\section{Details of the calculation}
\label{sec:calc}

We now proceed to push the leading order calculation of the last section
to the next order in HTL perturbation theory.

\subsection{Formalism and diagrams}
\label{sec:formalism}

The real-time correlator \Ref{correl1} can be expressed 
in terms of correlators of fields ordered along the Schwinger-Keldysh
contour \cite{Keldysh,ChouSuHaoYu}.
We find it convenient to use the Keldysh or $ra$ basis (where one works
in terms of the contour averaged or $r$ and contour differenced or $a$
fields rather than the fields on the upper $1$ and lower $2$ contours).
Throughout we will be using a graphical notation for the propagators
of this formalism, summarized in Table\ \ref{table:ra}.
We draw retarded ($ra$) and advanced ($ar$)
propagators with an arrow on them, which points in the direction
of the flow of time (thus towards the $r$ index.)  We draw 
the $rr$ propagators with a double cut in the middle of it;
there exists no $aa$ propagator in this formalism.
If one thinks of an $rr$ propagator as carrying two outgoing arrows,
leaving away from the cut, then an $r$ field at a vertex will have
an incoming arrow on it and an $a$ field will have an
outgoing arrow, leaving the vertex.
Thus the $ra$ assignments of the fields at the vertices
can be readily recovered from our notation.
The nonzero (tree-level) vertices in the Keldysh basis carry odd numbers
of $a$ indices, thus have an odd number of arrows leaving them
(but arbitrarily many $r$ indices, or incoming arrows.) 
Interaction vertices with one $a$ index are precisely the same as
the usual ones given in textbooks on
quantum field theory at zero temperature,
and those having three $a$ indices are smaller by a factor $\frac14$.
This notation is the same as in \cite{htlpaper}.

\begin{TABLE} {
\begin{tabular}{|c|c|c|} \hline
Symbol & Notation & Expression for a free scalar field \\ \hline
\putbox{0.10}{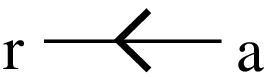}
& $G_{ra}(P)\equiv G_R(P)$ &
$\displaystyle \frac{-i}{P^2-i\epsilon p^0}$ \\ \hline
\putbox{0.10}{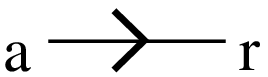}
& $G_{ar}(P)\equiv G_A(P)$ &
$\displaystyle \frac{-i}{P^2+i\epsilon p^0}$ \\ \hline
\putbox{0.10}{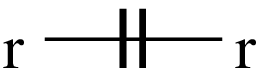}
& $G_{rr}(P)$ &
$\displaystyle 2\pi\delta(P^2)\,\left(\frac12 + n_B(|p^0|)\right)$ \\ \hline
\end{tabular}
\label{table:ra}
\caption{Graphical notation for real-time propagators in the Keldysh
basis, and their expression for a free scalar field.  In all cases
the momentum $P$ flows from right to left. }
} \end{TABLE}

Diagrammatic rules that generate the HTL effective vertices
have recently been
worked out for real time field theory in this basis \cite{htlpaper}; we
use the results and notation of that work, which are summarized
in Fig.~\ref{fig:htlrules}.
In the HTL limit the hard degrees of freedom behave
like classical point-like particles, which we draw as solid lines.
The vertices (a) and (c) and the eikonal propagator (d) of the figure
together describe
the generation and propagation of disturbances of the hard particles'
distribution functions in a background gauge field,
and the two-point vertex (b) describes how these disturbances
source gauge fields.
These effects depend on the four-velocity  $v^\mu=(1,\bv)$, $\bv^2=1$,
of the hard particle,
which has to be averaged over for every connected solid line that
appears in a diagram; a factor of $m_D^2/T$ must also be included.
The $rr$ propagator (e) describes statistical fluctuations in the
number density of the charges, and enters precisely once
in the calculation of HTL amplitudes with two external $a$ gluons.
The only HTL three point vertex that exists has the $ra$ assignment
shown in (c); there exist no HTL amplitudes with more than two external
$a$ gluons.
Diagrams containing self-energy insertions
on gluon propagators must be discarded, since we are already using
the HTL-resummed gluon propagators \Ref{HTLprop}.
The application of these rules reproduces calculations in
classical Yang-Mills plasmas
with point-like (nonabelian) charges; more details can be found
in \cite{htlpaper}.

\begin{FIGURE}[t] {
\renewcommand{\u}{\unitlength}
\begin{picture}(350,105)(0,10)
\put(0,45){\begin{picture}(100,100)
    \put(30,15){\includegraphics[width=60\u,height=25\u]{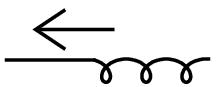}}
    \put(63,29){$P$}
    \put(83,29){$\mu$}
    \put(0,20){(a)}
    \put(25,18){a}
    \put(91,18){r}
    \put(105,20){$= ip^0\,v^\mu$}
\end{picture} }
\put(0,15){\begin{picture}(100,100)
    \put(30,15){\includegraphics[width=60\u,height=12\u]{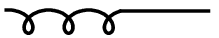}}
    \put(32,29){$\mu$}
    \put(0,20){(b)}
    \put(25,18){a}
    \put(91,18){r}
    \put(105,20){$= iv^\mu$}
\end{picture} }
\put(200,60){\begin{picture}(100,100)
    \put(30,20){\includegraphics[width=50\u,height=30\u]{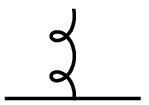}}
    \put(60,43){$\mu,a$}
    \put(33,25){$c$}
    \put(69,25){$b$}
    \put(0,20){(c)}
    \put(25,18){a}
    \put(80,18){r}
    \put(52,50){r}
    \put(90,20){$= -v^\mu f^{abc}$}
\end{picture} }
\put(200,30){\begin{picture}(100,100)
    \put(30,20){\includegraphics[width=50\u,height=12\u]{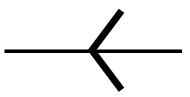}}
    \put(0,20){(d)}
    \put(25,18){a}
    \put(80,18){r}
    \put(63,29){$P$}
    \put(90,20){$= \displaystyle \frac{-i}{v\cdot P^-}$}
\end{picture} }
\put(200,0){\begin{picture}(100,100)
    \put(30,20){\includegraphics[width=50\u,height=12\u]{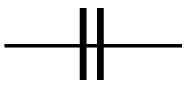}}
    \put(0,20){(e)}
    \put(25,18){r}
    \put(80,18){r}
    \put(63,29){$P$}
    \put(90,20){$= 2\pi\delta(v\cdot P)$}
\end{picture} }
\end{picture} 
\label{fig:htlrules}
\caption{Feynman rules for the HTL theory, with
$ra$ indices explicitly shown: (a)-(c) give interaction vertices
and (d)-(e) give effective propagators for classical particles.
All two-point functions are proportional
to the identity in color space, $\delta^{ab}$, not explicitly shown.
A factor $(m_D^2/T)$ plus an integration $\int\dv$ over the four-velocity
$v^\mu$ must be assigned to every disjoint solid line appearing in a diagram.}
} \end{FIGURE}

In the Keldysh basis, correlators involving Keldysh $a$ fields
with soft momenta $p\sim gT$ are systematically down by powers of $g$,
relative to similar correlators with Keldysh $r$ fields (see
e.g. \cite{htlpaper}), implying that at NLO all fields entering \Ref{correl1}
can be taken to be Keldysh $r$ fields.
This means that operator ordering issues actually are subleading,
and that at NLO the Wilson lines in \Ref{correl1}
can be traded for a single adjoint Wilson line.

It is convenient to write the electric field operators
as $E^i=-\partial^i A^0-D_t A^i$.  Using the equation of motion for a
Wilson line, $D_t W(t;0) = 0$, we can then express
\be
W(t;0)^\dagger \left(D_t A^{ia}(t)\, t^a_{\sss h}\right) W(t;0)
= \frac{d}{dt} \left( W(t;0)^\dagger \;  A^{ia}(t)\, t^a_{\sss h} \; W(t;0)
  \right) \, ,
\ee
which contributes a total derivative to \Eq{correl1} and can be
dropped.%
\footnote{This is true in any gauge in which the
propagators show no divergent or pathological behavior in the
zero-frequency limit. These includes the covariant or Coulomb gauges,
but not the temporal axial gauge.}
Therefore the $E^i$ can be replaced with $-\partial^i A^0$,
and the desired correlation function becomes
\be
\kappa = \frac{\ch g^4}{3d_A}\int_{-\infty}^\infty dt\,
\la \partial^i A^{0\,a}(t)
\left[{\cal P} e^{[\int_0^t dt' A^0(t'),\cdot]}\right]_{ab} 
     \partial^i A^{0\,b}(0)
\ra  \,. 
\label{kappacorrel}
\ee

\begin{FIGURE}[t] {
\renewcommand{\u}{\unitlength}
\begin{picture}(380,100)
\put(0,15){\includegraphics[width=110\u,height=80\u]{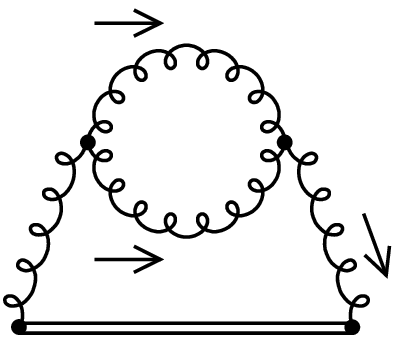}}
\put(45,0){(A)}
\put(51,30){$Q,b$}
\put(51,90){$R,c$}
\put(93,50){$P,a$}
\put(35,66){$\nu$}
\put(35,54){$\mu$}
\put(58,66){$\nu'$}
\put(58,54){$\mu'$}
\put(205,0){\begin{picture}(100,100)
  \put(0,15){\includegraphics[width=90\u,height=80\u]{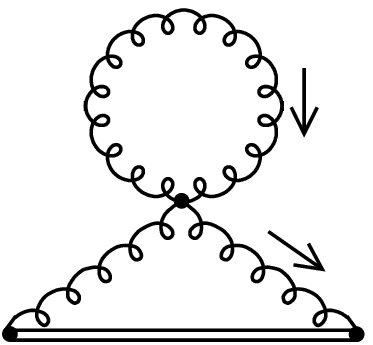}}
  \put(35,0){(C)}
  \put(70,44){$P$}
  \put(78,70){$Q$}
  \put(34,60){$\mu\,\,\,\,\,\,\nu$}
 \end{picture}}
\put(120,0){\begin{picture}(100,100)
  \put(0,15){\includegraphics[width=80\u,height=50\u]{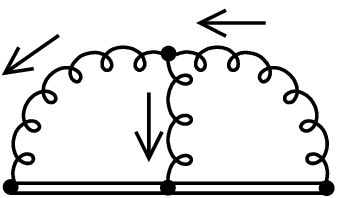}}
  \put(32,0){(B)}
  \put(15,58){$Q,a$}
  \put(12,26){$R,b$}
  \put(64,56){$P,c$}
 \end{picture}}
\put(302,15){\includegraphics[width=80\u,height=50\u]{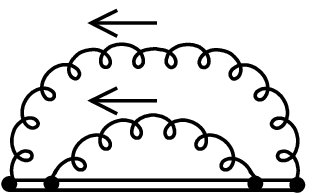}}
\put(335,0){(D)}
\put(345,39){$Q$}
\put(345,60){$P$}
\end{picture} 
\caption{The Feynman diagrams that contribute to $\kappa^{\rm NLO}$, with
assignments of momenta, Lorentz and color indices.
All propagators are soft and HTL-resummed,
and all interaction vertices include HTL corrections; the arrows
denote only the momentum flow.
The Lorentz indices of the gluons which connect to the heavy quark
(shown as the double line) are all ``0''.}
\label{fig:diagrams}
} \end{FIGURE}

Expanding the Wilson line gives a series of correlators of $A^0$
fields.  The diagrams we need at NLO are shown in
Fig.~\ref{fig:diagrams}.%
\footnote{%
     There is also a diagram which looks like (D) but with crossed gluon
     lines.  It vanishes by rotational invariance.%
    }
These diagrams are to be evaluated within the
HTL effective theory, meaning that all propagators are HTL resummed and
all vertices include HTL vertices--except for the vertices on the Wilson
line, shown as the double line in the diagram.
Naively there could be two more diagrams, corresponding to (A) and (C)
but with fermions rather than gauge bosons in the loops; but these are
suppressed by at least one factor of $n_{\sss f}/n_{\sss b} \sim g$
relative to the indicated diagrams and can be neglected at NLO.

\subsection{Real part of self-energy diagrams (A) and (C)}
\label{sec:re}

Diagrams (A) and (C) of Fig. \ref{fig:diagrams}
correspond to NLO self-energy corrections
to the soft zero-frequency longitudinal gluon propagator,
and can be decomposed into real and imaginary parts.
We begin with the real part of the self-energy. Since it is needed
only at zero frequency,
it can be most convenient evaluated within the imaginary time
formalism, in which the frequency integrals are replaced by discrete sums over
the (imaginary) Matsubara frequencies $\omega_n=2\pi nT$, $n$ integer \cite{kapusta}.
In this formalism, no analytic continuation of any kind is required
at \emph{zero} external frequency,
and we can directly analyze the discrete
sum over the Matsubara frequencies.

Because the HTL effective vertices vanish when all of their external
frequencies are zero, and are subleading by $g^2$
(and at any rate, inappropriate) when one of their external momenta
carries a nonzero Matsubara frequency $|\omega_n|\gsim T$,
the diagrams involving HTL vertices
do not contribute at NLO.
Similar cancellations occur for the transverse-transverse
contribution with tree interaction vertices, because
the relevant interaction vertex vanishes when all frequencies are zero,
and the contribution of nonzero Matsubara frequencies only
receives $\OO(g^2)$ corrections
(the presence of the hard frequency scale $\omega_n$ in the loop propagators
ensures that the self-energy corrections on the loop propagators
are down by $\OO(g^2)$, and that the $p$ dependence of the integral
over spatial momenta can be expanded
into integer powers of $p^2/T^2$ when $p\ll T\lsim \omega_n$.)
The diagrams with topology (C), involving four-point vertices,
similarly do not contribute: the one with an HTL vertex is irrelevant
at zero external frequency,
and when the vertex is a tree vertex
the propagator in the loop must be purely transverse (in the strict
Coulomb gauge)
since there is no interaction vertex involving only $A^0$ fields;
but the HTL correction to this propagator at zero frequency vanishes.
These simplifications, which
are specific to the zero-frequency retarded self-energy
(and to a lesser extent, to our use of strict Coulomb gauge),
are perhaps best understood
in terms of the dimensionally reduced effective
theory \cite{dimred}.

We are thus left to evaluate the transverse-longitudinal loop
with tree-level interaction vertices and HTL-resummed propagators.
Only the contribution from the zero Matsubara frequency is needed,
\bea \delta \Pi_R^L(p) &=&  -g^2 N_c T \int_q 4p^2
\left(1-\frac{(p\cdot q)^2}{p^2q^2}\right)
\frac{1}{q^2} \frac{1}{r^2+m_D^2}
\nl &=& -\frac{g^2 N_c T}{2\pi} \left[ m_D + \frac{p^2-m_D^2}{p}
  \tan^{-1}\left(\frac{p}{m_D}\right)\right]\,, \label{dummyre}
\eea
where $\int_q$ is shorthand for $\int \frac{d^3 q}{(2\pi)^3}$
and where the arctangent takes values in $[0,\frac{\pi}{2}]$.
The same result could also have been obtained within the
real-time formalism, albeit with somewhat more work.
This $\OO(g)$ correction to the real part of the longitudinal gluon self-energy
induces an $\OO(g)$ correction to $\kappa$,
\be \delta \kappa_{\rm Re} =
\frac{C_Hg^2}{3} \int_p p^2 \Pi^{>\,00}(p) \left[
\frac{1}{(p^2+m_D^2+\delta\Pi_R^L(p))^2}-\frac{1}{(p^2+m_D^2)^2}\right]
\ee
which, using the HTL approximation $\Pi^{>\,00}(p)=m_D^2\pi T/p$,
yields a contribution to the dimensionless coefficient $C$ from \Ref{defkappa}:
\bea C_{(A),{\rm Re}} &=& 6\pi \int_p \frac{p}{(1+p^2)^3}\left[1 +
\frac{p^2-1}{p} \sin^{-1}\left(\frac{p}{\sqrt{1+p^2}}\right)
\right] \nl
&=& \frac{3}{2\pi}\left( 1+ \frac{\pi^2}{16} \right) \simeq 0.77198914\ldots
\eea
where we have rescaled $p$ to $p/\mD$.  Note that the integral is both
IR and UV safe.

\subsection{Self-energy (A): imaginary part}
\label{sec:ima}

\subsubsection{Overview of the calculation}

The imaginary part of the gluon
self-energy diagram (A) is probably the most technically
challenging part of this calculation.
Instead of calculating the ($\OO(p^0)$ term of the)
imaginary part of the retarded self-energy,
we find it more convenient to calculate directly the Wightman self-energy
$\Pi^{>\,00}(P)=2(1+n_B(p^0))\Im \Pi_R^{00}$,
which can be evaluated directly at zero frequency.
There exists a finite temperature cutting rule, analogous to the
familiar zero-temperature Cutkowski rule, which expresses
this function in terms of a sum over diagrams that are
divided into two parts by one cut \cite{weldon83,pisarski87}%
\footnote{Although the rules given by these authors deal with
the imaginary part of the retarded self-energy, rather than the Wightman
self-energy, the rule we use follows from the latter by a
straightforward application of fluctuation-dissipation relations
(KMS conditions).
}.
The propagators traversed by the cut are to be evaluated as
Wightman propagators, $G^>(P)\equiv (G_R(P)-G_A(P))(1\pm n(p^0))$,
for bosons and fermions respectively,
and the ``amplitudes'' on each side of the cut are to be evaluated as
the fully-retarded amplitudes of the real-time formalism,
the retardation (e.g. time flow) being taken to be away from the cut, toward the
external legs.
In terms of the Keldysh $ra$ basis, this means that all cut propagators
attach to the neighboring vertices like Keldysh $r$ fields, and
external legs of the self-energy diagram should be considered as
carrying Keldysh $a$ indices.
These fully-retarded amplitudes are the simplest analytic continuation
of the imaginary-time amplitudes \cite{Evans} (they are obtained by continuing
all but one of the external momenta from the upper-half complex
frequency plane.)
A direct proof of the rule we use, within the real-time formalism,
has also been given using the $R/A$ formalism
\cite{Guerin} (see also \cite{gelis}, section 3.6.);
the cutting rule presented there is the same as the one we use,
since the fully retarded amplitudes of the $ra$ and $R/A$
formalisms are the same%
\footnote{Since the most complicated self-energy diagram we need to evaluate
contains ``only'' three loops, it is also possible to give a direct
proof of the cutting rule in our case, starting from the standard rules of the
Schwinger-Keldysh $ra$ formalism applied to the calculation of the $aa$ self-energy
(which is the average of the two Wightman self-energies.)
We have checked this; although somewhat long the proof is a succession of
simple manipulations, which only involve the addition or subtraction
of suitable closed loops of retarded propagators to the diagrams
(such closed loops in a diagram evaluate to zero.)
}.

The four distinct types of cuts which contribute to the self-energy
diagrams with two HTL effective vertices are depicted in
Fig.~\ref{fig:cuta}.  As just mentioned, all cut propagators are Wightman
propagators, and they attach to the neighboring vertices like
Keldysh $r$ fields.
When soft gluon propagators are traversed by the cut, the small $P$ approximation
$G^>(P)\approx G_{rr}(P)$ can be used.
When HTL vertices are traversed by the cut,
two hard propagators are put on-shell;
as discussed in more detail in \cite{htlpaper}, the corresponding
amplitudes are precisely given by the real-time
HTL amplitudes having two external Keldysh $a$ indices.
Physically, these amplitudes are obtained by making the eikonal
approximation in all propagators and vertices entering in the hard
loop; the interested reader may readily verify that this
reproduces the HTL amplitudes with two Keldysh $a$ indices as given
by the rules of section \ref{sec:formalism}.

The cuts of type (i)-(iii) share a feature which is very pleasing from
the viewpoint of
their numerical evaluation: they are given by expressions
supported on the spectral weights of the soft gluon propagators.
In other words they split into pole-pole, pole-cut and cut-cut parts,
according to whether the momenta $Q$ and $R$ are restricted
to lie on the position of a plasmon pole,
or to lie within the space-like region (``Landau cut'').
This should be obvious from Fig.~\ref{fig:cuta}.
In contrast, cut (iv), which represents a (two-loop!) virtual correction
to the tree processes considered in section \ref{sec:physics},
might be expected to induce additional complications
since it leaves essentially unconstrained the gluon momenta that appear
in it.
However, somewhat to our surprise, under the special circumstance
$p^0=0$ we were able to bring this contribution
into a form manifestly supported within the cut-cut region.
The relevant manipulations are described
below in greater detail.
As far as we know, this additional difficulty did not
show up in previous HTL calculations,
such as Braaten and Pisarski's pioneering evaluation of the
gluon damping rate \cite{braatendamping}, or Braaten, Pisarski
and Yuan's calculation of soft dilepton production \cite{braatenyuan},
cut (iv) being kinematically forbidden in these cases due to the
the external momentum being time-like.

\begin{FIGURE} {
\includegraphics[width=15cm,height=2.1cm]{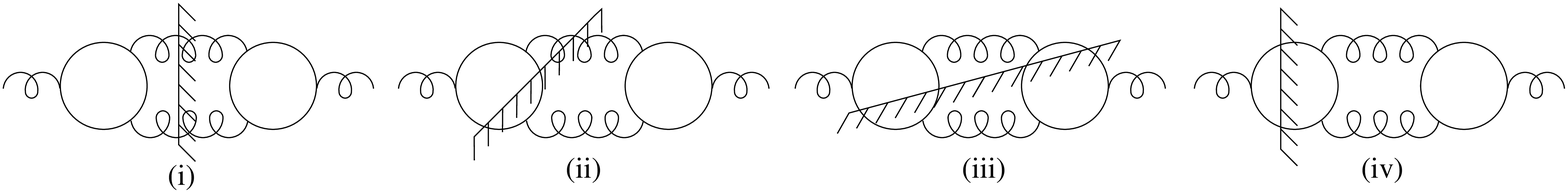}
\caption{The distinct cuts which can go through the self-energy
diagram (A) of Fig. \ref{fig:diagrams} with two HTL effective vertices
(drawn as loops), as explained in the text.
Solid lines denote hard propagators and the two
gluon propagators carry soft momenta.}
\label{fig:cuta}
} \end{FIGURE}

\subsubsection{Evaluation of the cuts}
\label{sec:cuta}

In Fig.~\ref{fig:blocks} we give explicit expressions for the HTL effective
vertices entering Fig.~\ref{fig:cuta} (i)-(iii), in terms of certain
functions,
\bea
M^{\mu\nu}(Q,R) &\equiv& \int \dv \frac{v^\mu v^\nu}
{v\cdot Q^- v\cdot R^-} \label{defM}\,, \\
K^{\mu\nu}(Q,P) &\equiv& \int \dv i\pi\delta(v\cdot Q)
\frac{v^\mu v^\nu}{v\cdot P^-}\,,  \label{defK} \\
V^{\mu\nu}(Q,R) &=& \frac{-1}{m_D^2} \left[
2q^0 \eta^{\mu\nu} +(R+P)^\mu \eta^{\nu0}
-(Q+P)^\nu\eta^{\mu0}\right]\,. \label{defV}
\eea
which are related to the fully retarded HTL three-point vertex, its
discontinuities,
and to the tree vertex, respectively.  The Keldysh indices
that appear on these effective vertices, for the cuts (i)-(iii),
are completely determined by the cutting rule we use.

\begin{FIGURE} {
\renewcommand{\u}{\unitlength}
\begin{picture}(320,185)(0,0)  
\put(0,145){\begin{picture}(100,50)
    \put(30,5){\includegraphics[width=50\u,height=30\u]{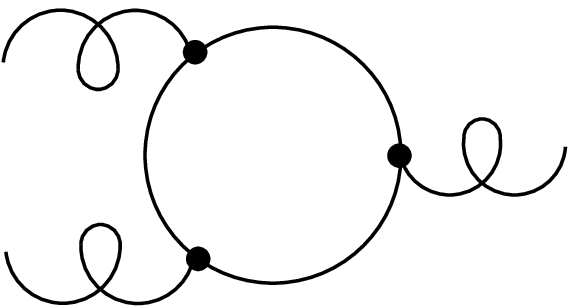}}
    \put(0,15){(1)}
    \put(43,36){r}
    \put(43,-2){r}
    \put(68,25){a}
    \put(90,23) {
      $ \displaystyle = m_D^2 g \,i^2 (-i)^2 f^{a'bc} \int_v \frac{v^{\mu'}v^{\nu'}}{v\cdot P^-}
      \left[ \frac{q^0}{v\cdot Q^-} - \frac{r^0}{v\cdot R^-}\right]$ }
    \put(110,0) {
      $ \equiv m_D^2 g f^{a'bc} \times q^0 M^{\mu'\nu'}(Q,R) $ }
\end{picture} }

\put(0,95){\begin{picture}(100,50)
    \put(30,5) {\includegraphics[width=50\u,height=30\u]{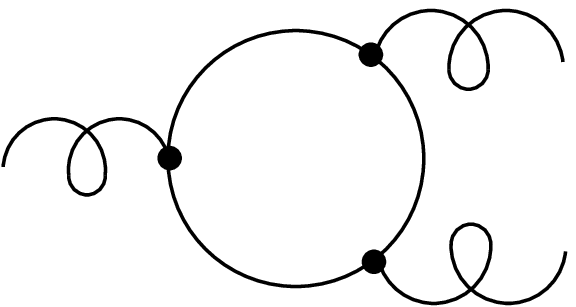}}
    \put(0,15){(2)}
    \put(62,36){r}
    \put(62,-2){r}
    \put(38,25){a}
    \put(90,18){ 
      $ = m_D^2 g f^{abc} \times -q^0 M^{\mu\nu}(Q,R)^* $}
\end{picture} }

\put(0,50){\begin{picture}(100,50)
    \put(30,5) {\includegraphics[width=50\u,height=33\u]{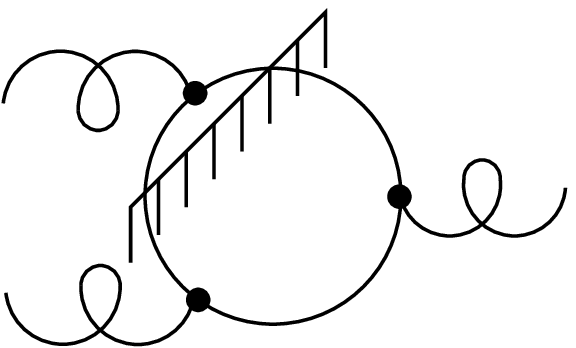}}
    \put(0,15){(3)}
    \put(43,35){a}
    \put(43,-2){r}
    \put(68,25){a}
    \put(90,23) {
      $ \displaystyle = m_D^2 gT\, i^3 f^{a'bc}\int_v \frac{v^{\mu'}v^{\nu'}}{v\cdot P^-}
      2\pi\delta(v\cdot R) $ }
    \put(110,0){ 
      $ \equiv m_D^2 g f^{a'bc} \times -2K^{\mu'\nu'}(R,P) $ }
\end{picture} }

\put(0,0){\begin{picture}(100,50)
    \put(30,5) {\includegraphics[width=50\u,height=30\u]{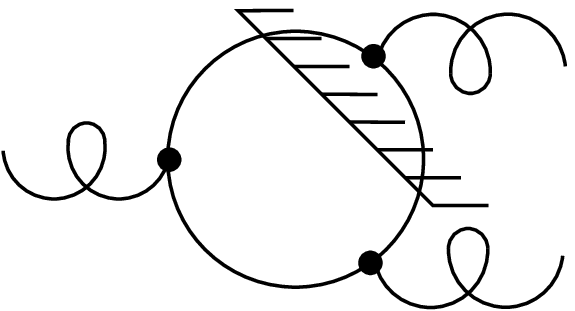}}
    \put(0,15){(4)}
    \put(62,36){a}
    \put(62,-2){r}
    \put(38,25){a}
    \put(90,18){
      $ = m_D^2 g f^{abc} \times 2K^{\mu\nu}(Q,P)^* $ }
\end{picture} }
\end{picture} 
\caption{The HTL effective vertices that appear
in Fig.\ \ref{fig:cuta} (i)-(iii), with $ra$ indices shown.
The momenta, Lorentz and color indices are as suggested by the position
of these objects in Figs.\ \ref{fig:diagrams}, \ref{fig:cuta}.}
\label{fig:blocks}
} \end{FIGURE}

With these basic building blocks in hands, the evaluation
of the cuts of type (i)-(iii) is relatively
straightforward (our normalization is $\Pi^{>\,00}(P)=2\Im
\Pi_R^{00}(P)(T/p^0)$ with $\Pi_R^{00}$ as in \Ref{HTLprop}):
\bea
\frac{\Pi^{>\,00}_{(A),{\rm (i)}}(p)}{N_c m_D^4 g^2T^2/2} &=&
\int_Q M_{\mu\nu}(Q,R) M_{\mu'\nu'}(Q,R)^*
\,\rho^{\mu\mu'}(Q) \rho^{\nu\nu'}(-R) \,,
\nl
\frac{\Pi^{>\,00}_{(A),{\rm (ii)}}(p)}{N_c m_D^4 g^2T^2/2} &=&
\int_Q  \left[ -2M_{\mu\nu}(Q,R)K_{\mu'\nu'}(Q,P)^*
\, G_R^{\mu\mu'}(Q) \rho^{\nu\nu'}(-R) \right.
\nl
&& \hspace{2cm} 
\left. +2K_{\mu\nu}(Q,P) M_{\mu'\nu'}(Q,R)^*
\, G_A^{\mu\mu'}(Q) \rho^{\nu\nu'}(-R) \right]\nl
&& \hspace{1cm} + (Q\leftrightarrow R)\,,
\nl
\frac{\Pi^{>\,00}_{(A),{\rm (iii)}}(p)}{N_c m_D^4 g^2T^2/2} &=&
\int_Q 4K_{\mu\nu}(R,P)K_{\mu\nu'}^*(Q,P)\, G_R^{\mu\mu'}(Q) G_A^{\nu\nu'}(R)
\nl && \hspace{1cm} + (Q\leftrightarrow R) \label{cutabc}\,,
\eea
all of which are manifestly real.  We are using the approximation
$G_{rr}(Q)\approx \rho(Q)/q^0$, with
$\rho\equiv (G_R-G_A)$ being the spectral density.
The contribution from cut (i) is also manifestly positive, as expected
from its rather obvious interpretation as the square of a one-loop
amplitude (although at this stage it might not be obvious that the
sum over Lorentz indices yields a sum of
positive terms; this is confirmed in section \ref{sec:lorentz}.)
In \Ref{cutabc} the contribution from diagrams involving the tree
interaction vertices is not shown explicitly; it can be recovered
by the simple
substitution, for each appearance of the fully retarded HTL vertex
$M_{\mu\nu}(Q,R)$ or its complex conjugate,
\be q^0M_{\mu\nu}(Q,R)\to q^0M_{\mu\nu}(Q,R) + V_{\mu\nu}(Q,R) \,. \label{substbare}
\ee

\begin{FIGURE} {
\putbox{0.30}{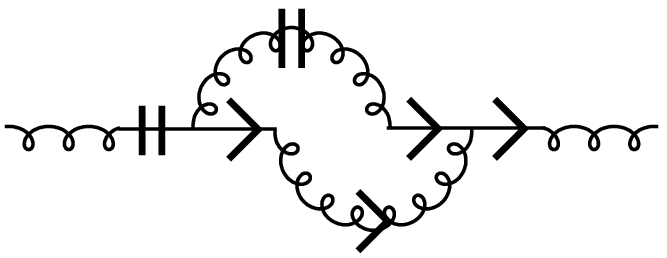}
\hfill
\putbox{0.30}{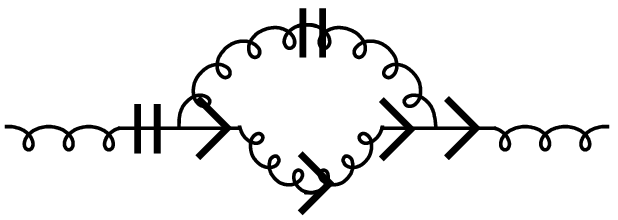}
\hfill
\putbox{0.30}{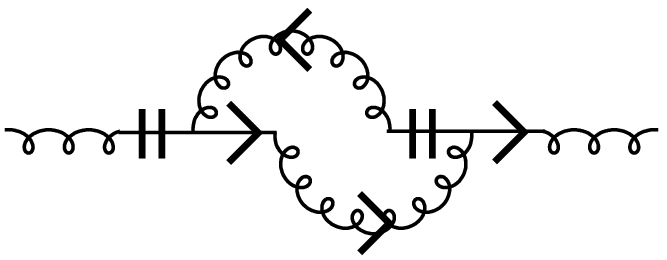}
\hfill
\putbox{0.25}{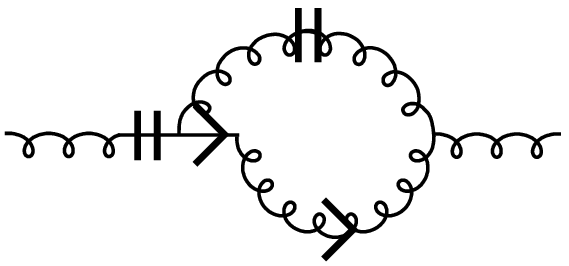}
\caption[NLO diagrams]
{ \label{fig:cutd}
Zoom on the $ra$ structure of the
propagators appearing in the HTL diagrams
contributing to cut (iv) of Fig.~\ref{fig:cuta}.}
} \end{FIGURE}

Evaluating the cuts of type (iv) poses some additional
difficulty, as mentioned above.
The HTL diagrams contributing to it are depicted in
Fig.~\ref{fig:cuta}.  The leftmost solid line in these diagrams
can be identified with the cut hard loop on the left-hand side of
diagram (iv) in Fig.~\ref{fig:cuta},
collapsed to a one-dimensional line \cite{htlpaper}:
the cut ($rr$) HTL propagator replaces the two cut hard propagators in
Fig.~\ref{fig:cuta} (iv), and the retarded HTL propagator on its right
replaces the (eikonalized) third propagator of this hard loop.
The reason why there appears exactly one cut gluon, or HTL propagator,
in the rest of the diagram is because
the object which stands on the right-hand side of the
explicit cut in Fig.~\ref{fig:cuta} (iv) is a one-loop
retarded amplitude (in the HTL theory.)
The direct evaluation of these HTL diagrams yields:
\bea \Pi^{>\,00}_{(A),{\rm (iv)}}(P)
 &=& +2\Re N_c m_D^4 g^2 T^2  \int_Q\, 2K_{\mu\nu}(P,Q)
\nl
&& \hspace{1cm} \times \left\{ \begin{array}{l} \D
M_{\mu'\nu'}(Q,P) G_R^{\mu\mu'}(Q) \rho^{\nu\nu'}(R) \frac{q^0}{r^0}
\\ \D
-M_{\mu'\nu'}(R,P) G_R^{\mu\mu'}(Q) \rho^{\nu\nu'}(R)\frac{r^0}{r^0}
\\
-2K_{\mu'\nu'}(R,P) G_R^{\mu\mu'}(Q) G_A^{\nu\nu'}(R)
+V_{\mu'\nu'}(Q,P) G_R^{\mu\mu'}(Q) \rho^{\nu\nu'}(R) \frac{1}{r^0}\,.
\end{array} \right. \label{cutivI}
\eea
Using standard methods of contour integration that make use
of the analyticity of the prefactor $K_{\mu\nu}(P,Q)$
in the upper half $q^0$ plane, this expression can be rewritten
more compactly. Specifically, we introduce a small
$i\epsilon$ prescription $1/r^0\to 1/(r^0-i\epsilon)$ in all places
this appears.
This is equivalent
to displacing the contour of $r^0$ integration slightly below the real
axis (or, equivalently, the contour of $q^0$ integration slightly
above the real axis),
and does not change the final answer since the numerator
vanishes at $r^0=0$, $\rho(R)$ being an odd function
of $r^0$.  However, the introduction of this prescription
makes it possible to decompose $\rho(R)$ into $(G_R(R)-G_A(R))$,
the integral of each term remaining well-defined.
Decomposing also $2K(R,P)$ into $(M(R,P)+M(-R,P))$,
and dropping all terms which are analytic in the upper-half
$q^0$ plane, one obtains:
\bea
\Pi^{>\,00}_{(A),{\rm (iv)}}(P)
 &=& +2\Re N_c m_D^4 g^2 T^2  \int_Q\, 2K_{\mu\nu}(P,Q)
G_R^{\mu\mu'}(Q) G_R^{\nu\nu'}(R) \nl && \times \left\{
\frac{q^0M_{\mu'\nu'}(Q,P)-r^0M_{\mu'\nu'}(R,P)}{r^0-i\epsilon}
+ \frac{V_{\mu'\nu'}(Q,P) }{r^0-i\epsilon}
\right\}\,. \label{cutivII}
\eea
Although \Ref{cutivII} is valid for arbitrary $p^0$,
a great simplification occurs when $p^0=0$: the denominator
$1/(r^0-i\epsilon)$ is antisymmetric
under $Q\leftrightarrow R$, modulo its $i\epsilon$ prescription.
By enforcing symmetry of the integrand under $(Q\leftrightarrow R)$
we can thus trade the prefactor $2K(P,Q)$ (whose support extends
to $q^0\to \infty$, a nuisance for numerical work)
into the better-behaved combination
$K(P,Q)+K(P,R)=2\Re K(Q,R)$, whose support lies entirely
within the region of spacelike $Q$ and $R$, $|q^0|<{\rm min}(q,r)$.
Due to the explicit factor of $q^0$ present in the numerator,
the $i\epsilon$ prescription in the denominator of the HTL terms
(the one involving $M_{\mu'\nu'}$) actually
is unimportant and can be discarded,
allowing $(Q\leftrightarrow R)$ symmetry to be enforced at no cost.
However, no such factor of $q^0$ multiplies the term involving
the tree vertex $V_{\mu'\nu'}$,
and enforcing the $(Q\leftrightarrow R)$ symmetry in it gives rise
to an additional contribution proportional to $\delta(q^0)$,
coming from the mismatch of $i\epsilon$ prescriptions.
Thus at $p^0=0$ \Ref{cutivII} becomes:
\bea \hspace{-0.8cm}
\frac{\Pi^{>\,00}_{(A),{\rm (iv)}}(p)}{N_c m_D^4 g^2T^2/2}
&=& -8\Re\int_Q \left[\Re K_{\mu\nu}(Q,R)\right]
G_R^{\mu\mu'}(Q)G_R^{\nu\nu'}(R)
\,\left[M_{\mu'\nu'}(Q,R) +\frac{V_{\mu\nu}(Q,R)}{q^0}\right]
\nl &&
-4 \int_q V_{i0}(q,r) G_R^{T}(q) G_R^{00}(r)
\left(\delta^{ij}-\frac{q^iq^j}{q^2}\right)
\Im \left[ K_{j0}(p,q)-K_{j0}(p,r)\right]\,.
\label{cutivIII}
\eea
We remark that the apparent singularity at $q^0$, in the term involving
the tree vertex on the first line, is illusory (this
is why we dropped the $i\epsilon$ prescription in it.)
Indeed, since the tree vertex with three $A^0$ fields is
identically zero, and the tree vertex between one $A^0$ and two
transverse gauge fields is explicitly proportional to $q^0$,
a singularity could only happen when one gluon is transverse
and the other one is longitudinal. However, in that case
the prefactor $\Re K_{i0}(Q,R)$ turns out to be
explicitly proportional to $q^0$. In terms of the formulae for the
Lorentz algebra given
in the next section, the corresponding statement is that $X_{T-L}\propto
q^0$. Thus the first line of \Ref{cutivIII} is not sensitive
to the $q^0\to 0$ region. However, although the prefactor
$\Re K_{i0}(Q,R)=\frac12(K_{i0}(P,Q)+K_{i0}(P,R))$ vanishes at $q^0=0$,
$K_{i0}(P,Q)$ itself does not. This is the reason why one gets
a nonzero residue at $q^0=0$ in the transverse-longitudinal case
(and only in that case), as given on the second line of \Ref{cutivIII}.

To bring these expressions
into a form suitable for numerical evaluation, we find it convenient
to decompose all individual factors into their real and imaginary parts.
Doing so, the contributions \Ref{cutabc} and the first line
of \Ref{cutivIII} add up to ($\rho=2\Re G_R$):
\bea \frac{\Pi^{>\,00}_{(A)}(P)}{N_cm_D^4g^2 T^2/2}
&=& \int_Q \left\{ \begin{array}{l}
\rho(Q)\rho(-R)\,\left[(\Re M-\Re K)^2 - (\Im K_Q)^2 
 \right . \\ \hspace{2.5cm} \left.
- (\Im K_R)^2 +  (\Re K)^2\right]
\\
+4\Im G(Q) \rho(R)\left[ (\Re M -\Re K)\Im K_Q + \Re K \Im K_R\right] \\
+4\rho(Q) \Im G(R)\left[ (\Re M -\Re K)\Im K_R + \Re K \Im K_Q\right]
\\
+8\Im G(Q) \Im G(R) \left[ (\Re M -\Re K)\Re K \right.\\
\hspace{4cm}\left.- \Im K_Q \Im K_R\right].
\end{array} \right.
\label{theexpression1}
\eea
Here we have not explicitly written the contributions involving
tree vertices, which are recovered by the
simple substitution \Ref{substbare},
and we have dropped all Lorentz indices, which play no crucial role
here. We are using the abbreviations $M\equiv M(Q,R)$,
$K_Q\equiv K(Q,P)$ and $K_R\equiv K(R,P)$.
This expression incorporates a wealth of real and virtual
physical processes, as discussed in Subsection \ref{sec:physics}.

In addition to \Ref{theexpression1} we have the contribution from the
$q^0=0$ residue, given by the second line of \Ref{cutivIII}.
Using \Ref{defV} for $V_{i0}$ and performing the $v$ integration
we can make the latter more explicit:
\be \frac{\Pi^{>\,00}_{(A),q^0=0}(P)}{N_cm_D^4g^2 T^2/2} =
\frac{8\pi}{\mD^2 p} \int_q \frac{1}{q^2(r^2+m_D^2)}
\frac{p\cdot q}{q^2}
= \frac{1}{\mD p}\left[ \frac{\tan^{-1}\left(\frac{p}{\mD}\right)}{\mD p} -
\frac{1}{p^2+\mD^2}\right]\,.
\label{bizarreremainder}
\ee
Here we performed the $\bq$ integration
by first doing the integration over the angle
between $\bq$ and $\bp$, and then evaluating the integration over the
magnitude $q$ from its discontinuities at
its branch cuts at $q=\pm p + i[\mD,\infty)$.
This zero-frequency contribution to \Ref{cutivIII} may appear odd-looking,
compared to \Ref{theexpression}. 
However, what we regard
as truly remarkable, is the fact that the contribution from cut (iv)
(a two-loop virtual correction!),
\emph{could}, when $p^0=0$, be cast into a computer-friendly
form supported on the spectral weights of the gluon propagators.
The leftover piece \Ref{bizarreremainder}
seems to be the price to pay for this welcomed simplification.
It is not clear to the authors whether such a structure persists for
general spacelike $P$ with $p^0\neq 0$.

\subsubsection{The Lorentz structure}
\label{sec:lorentz}

We have to sum over the Lorentz indices in expressions of the form
\be M_{\mu\nu}(Q,R) G^{\mu\mu'}(Q)G^{\nu\nu'}(R)M_{\mu'\nu'}^*(Q,R)\,.
\label{lorentzsum0}
\ee
This is where our choice of strict Coulomb gauge becomes particularly
convenient.  First, in this gauge the retarded and cut propagators have 
the same Lorentz structure, see \Ref{HTLprop}.
Second, in this gauge the propagator decomposes into a longitudinal part
$G^{00}$ and two spatial, strictly transverse components,
$G_{ij} = G^T ( \epsilon_i \epsilon_j + \epsilon'_i \epsilon'_j )$.
We can choose one of these components, say, $\epsilon$,
to lie in the plane defined by $\q$ and $\r$ and the other, $\epsilon'$,
to be orthogonal to this plane.

These three components of the gauge propagator will give rise to four
structures in evaluating \Eq{lorentzsum0}; one contribution proportional
to $(G^{00})^2$, one contribution proportional to $G^{00} G^T$, and two
contributions proportional to $(G^T)^2$, one of which arises from the out
of plane and one from the in-plane polarization states.

The doubly longitudinal contribution to \Eq{lorentzsum0} is trivial; it 
is $|M^{00}|^2 G^{00}(Q) G^{00}(R)$.  Consider next the $G^{00}(R) G^T(Q)$
contribution.  To evaluate the contribution we need to study
$M_{i0}(Q,R)$, which, viewed as a vector, must involve a linear
combination of $q^i$ and
$r^i$. Since the coefficient of $q^i$ is annihilated by the $Q$ transverse
projector, only the coefficient of $r^i$ contributes to
\Ref{lorentzsum0}.  We can find this contribution by applying a
projector which removes the piece proportional to $q^i$:
\bea 
  M^{i0}(Q,R) &\equiv& r^i M_{T-L}(Q,R) + \mbox{ Terms proportional
  to } q^i\,, \nl
  M_{T-L}(Q,R) &=& \frac{1}{p^2q_\perp^2} (q^2 r^i - q\cdot r q^i)M^{i0}\,,
\eea
in which $q_\perp$ denotes the component of $q$ perpendicular to $p$,
$q_\perp^2p^2\equiv |q\times p|^2 = |q\times r|^2 = |r\times p|^2$.

Now consider the contributions where both propagators are transverse.
The function $M_{ij}$ vanishes when contracted against one in-plane and
one out-of-plane polarization vector, by parity invariance in the
out-of-plane direction.  Therefore there are two contributions, one
arising from the in-plane projection of $M_{ij}$ and one from the
out-of-plane projection of $M_{ij}$.  The out-of-plane projection
is
\be 
 M_{T-T,A}(Q,R) =
  \left( \delta^{ij} - \frac{r^2q^iq^j+q^2r^ir^j
         -q\cdot r q^ir^j - q\cdot r q^jr^i}{p^2 q_\perp^2}\right) 
         M^{ij} 
\label{dummyTTA}\,,
\ee
and the in-plane projection, obtained by dotting $M_{ij}$ against the
in-plane polarization operator for each propagator, is
\be 
M_{T-T,B}(Q,R) = \frac{q r}{p^2 q_\perp^2}
\left(r^i - \frac{q\cdot r}{q^2} q^i\right) 
\left(q^j - \frac{r\cdot q}{r^2} r^j\right) M^{ij} \,.
\ee

Using this procedure we find
\bea \Ref{lorentzsum0}&=& |M^{00}|^2 G^{00}(Q)G^{00}(R) +
2r^2q_\perp^2 |M_{T-L}(Q,R)|^2G^T(Q)G^{00}(R)
\nl && + |M_{T-T,A}(Q,R)|^2G^T(Q)G^T(R) +
|M_{T-T,B}(Q,R)|^2G^T(Q)G^T(R)\,. 
\label{dummylorentz}
\eea

We have evaluated the scalar functions entering \Ref{dummylorentz}
in terms of linear combinations of
$M^{00}(Q,R)$, 1, and two new functions $L(Q)$ and $L(R)$, with
momentum-dependent
coefficients (that are real and analytic functions of $q^0$.)
In a condensed notation the result can be written:
\bea \Ref{dummylorentz}
&=& \sum_i P_i\, G^{Q_i}(Q) G^{R_i}(R) M_{i}(Q,R) M_{i}(Q,R)^* \label{decomp1}\,, \\
 M_i(Q,R)&=&X_i M^{00}(Q,R) + Y_{Q_i} L(Q) + Y_{R_i} L(R) + Z_i \label{decomp2}\,,
\eea
with
\be L(Q)\equiv \int \dv\frac{1}{v\cdot Q^-}
\ee
and $M^{00}(Q,R)$ as defined in \Ref{defM}.
The sum over $i$ in \Ref{decomp1} covers the four cases L-L, T-L, T-T,A and T-T,B.
The momentum-dependent coefficients $X_i$, $Y_{Q_i}$, $Y_{R_i}$ and $Z$,
as well as the prefactors $P_i$ and choices of propagators, $G^{Q_i}$
and $G^{R_i}$, are tabulated in Table \Ref{tab:lorentz}. In this table
we have separated the contributions to the ``constant term'' $Z$ coming
from the HTL and tree vertices.
The various discontinuities of $M_{\mu\nu}$ which enter
\Ref{theexpression1} can be obtained from the discontinuities
of the basis functions entering \Ref{decomp2},
which are described in detail in Appendix \ref{app:fcts}.

\begin{TABLE} {
\begin{tabular}{|c|c|c|c|c|} \hline
Index $i$ & Prefactor $P_i$ & $G^{Q_i}$ & $G^{R_i}$ & $X_i$ \\ \hline
L-L & 1 & $G^{00}$ & $G^{00}$ & 1  \\ \hline
T-L & $2/q^2p^2q_\perp^2$ & $G^{T}$ & $G^{00}$ & $-q^0 p\cdot q$ \\ \hline
T-T,A & 1 & $G^{T}$ & $G^{T}$ & $1-q_0^2/q_\perp^2$ \\ \hline
T-T,B & $1/q_\perp^4 p^4 q^2 r^2$ & $G^{T}$ & $G^{T}$
& $-q_0^2p\cdot q p\cdot r$  \\ \hline
\end{tabular}
\\ \vspace{2mm} \\
\begin{tabular}{|c|c|c|c|c|}  \hline 
Index $i$ & $Y_{Q_i}$ & $Y_{R_i}$ & $Z_{i}^{({\rm HTL})}$ &
$Z_i^{({\rm tree})}$ \\ \hline
L-L & 0 & 0 & 0 & 0 \\ \hline
T-L & $q^2$ & $-q\cdot r$ & 0 & $2/q^0m_D^2$ \\ \hline
T-T,A & $q^0 p\cdot q/p^2q_\perp^2$ & $-q^0 p\cdot r/p^2q_\perp^2$ & 0 &
$-2/m_D^2$ \\ \hline
T-T,B & $q^0 q^2p\cdot r$ &
$-q^0 r^2p\cdot q$ & $p^2 q_\perp^2$ &
$2q\cdot r q_\perp^2 p^2/m_D^2$ \\ \hline  
\end{tabular}
\vspace{2mm}
\caption[The coefficients in the expansion \Ref{decomp1}, of the
sum over Lorentz indices.]
{The coefficients in the expansion \Ref{decomp1}.
The coefficient $Z_i$ appearing in the text is the sum of its
HTL and tree contributions $Z_i^{({\rm HTL})}$ and $Z_i^{({\rm tree})}$,
respectively. }
\label{tab:lorentz}
} \end{TABLE}

\subsubsection{Final expressions for diagram (A)}
\label{sec:finala}

The Wightman self-energy $\Pi^{>\,00}$ enters the heavy quark
diffusion coefficient $\kappa$ through:
\be \kappa_{(A)}= \frac{g^2C_H}{3}
\int \frac{d^3p}{(2\pi)^3} \frac{p^2}{(p^2+\mD^2)^2} \Pi^{>\,00}_{(A)}(p)\,.
\label{pintegration}
\ee
Substituting formula \Ref{theexpression1} into this,
upon rescaling variables by $\mD$ and scaling out the prefactor
from \Ref{defkappa} in order to obtain the dimensionless contribution to $C$,
we obtain, using the decomposition \Ref{decomp1} and the results
of Appendix \ref{app:fcts}:
\bea
\lefteqn{ C_{(A),{\rm main}} = 3\pi \int_p \frac{p^2}{(1+p^2)^2} \int_Q
\sum_i P_i }&&\nl && \times \left\{  \begin{array}{l}
\rho^{Q_i}(Q)\rho^{R_i}(-R)\,\left[(\Re M_i-\Re K_i)^2 - (\Im K_{Q_i})^2 
 \right . \\ \hspace{2.5cm} \left.
- (\Im K_{R_i})^2 +  (\Re K_i)^2\right]
\\
+4\Im G^{Q_i}(Q) \rho^{R_i}(R)\left[ (\Re M_i -\Re K_i)\Im K_{Q_i} + \Re K_i \Im K_{R_i}\right] \\
+4\rho^{Q_i}(Q) \Im G^{R_i}(R)\left[ (\Re M_i -\Re K_i)\Im K_{R_i} + \Re K_i \Im K_{Q_i}\right]
\\
+8\Im G^{Q_i}(Q) \Im G^{R_i}(R)\left[ (\Re M_i -\Re K_i)\Re K_i - \Im K_{Q_i} \Im K_{R_i}\right],
\end{array} \right. \label{theexpression}
\eea
in which:
\bea
\Re M_i - \Re K_i &\equiv&
-X_i\frac{
\tan^{-1} \left(\frac{p\sqrt{q_\perp^2-q_0^2}}{p\cdot q}\right) 
+\tan^{-1} \left(\frac{p\sqrt{q_\perp^2-q_0^2}}{p\cdot r}\right) 
-\frac{\pi}{2}}{p\sqrt{q_\perp^2-q_0^2}}
\nl && - Y_{Q_i} \frac{1}{2q} \ln\left(\frac{q+q^0}{q-q^0}\right)
   - Y_{R_i} \frac{1}{2r} \ln\left(\frac{r-q^0}{r+q^0}\right) + Z_i,
\nl
\Im K_{Q_i} &\equiv& Y_{Q_i} \frac{\pi}{2q}, \nl
\Im K_{R_i} &\equiv& Y_{R_i} \frac{\pi}{2r}, \nl  
\Re K_i &\equiv& -X_i \frac{\pi}{2p\sqrt{q_\perp^2-q_0^2}},
\label{refstuff1}
\eea
when $|q^0|<q_\perp$, and:
\bea
\Re M_i - \Re K_i &\equiv&
-X_i \frac{\ln\left(
\frac{|p\cdot q + \sqrt{q_0^2-q_\perp^2}|}
     {|p\cdot q - \sqrt{q_0^2-q_\perp^2}|}
\frac{|p\cdot r + \sqrt{q_0^2-q_\perp^2}|}
     {|p\cdot r - \sqrt{q_0^2-q_\perp^2}|} \right)
}{2p\sqrt{q_0^2-q_\perp^2}} \nl
&& - Y_{Q_i} \frac{1}{2q} \ln\left(\frac{|q^0+q|}{|q^0-q|}\right)
   - Y_{R_i} \frac{1}{2r} \ln\left(\frac{|q^0-r|}{|q^0+r|}\right) + Z_i,
\nl 
\Im K_{Q_i} &\equiv& X_i \frac{\pi\sgn(p\cdot
  q)}{2p\sqrt{q_0^2-q_\perp^2}}
+ Y_{Q_i} \frac{\pi}{2q}\theta(q^2-q_0^2),  \nl
\Im K_{R_i} &\equiv& -X_i \frac{\pi\sgn(p\cdot
  r)}{2p\sqrt{q_0^2-q_\perp^2}}
+ Y_{R_i} \frac{\pi}{2r}\theta(r^2-q_0^2), \nl  
\Re K_i &\equiv& 0,
\label{refstuff2}
\eea
when $|q^0|>q_\perp$.
The arctangents on the first line of \Ref{refstuff1}
take values in $[0,\pi]$. 
The coefficients $P_i$, $X_i$, $Y_{Q_i}$, $Y_{R_i}$ and $Z_i$, as well
as the choices of propagator (transverse or longitudinal),
for the different choices of $i$,
are listed in Table \ref{tab:lorentz}. Expressions for the
HTL-resummed retarded propagators $G_R$ and $\rho=G_R-G_A$ are
given in \Ref{HTLprop}.
The integrals depend only on one scale,
$\mD$, which we have scaled out and should be set to 1 whenever 
it shows up in the formulae. 
As was discussed earlier, the integrand naturally splits
into pole-pole, pole-cut and cut-cut contributions, according to whether
the energies of the $Q$ and $R$ propagators lie within the space-like
region (the Landau cut) or on the plasmon pole.

The integral \Ref{theexpression} is (linearly) divergent at large $q$,
due to the
transverse-transverse pole-pole contribution involving tree interaction
vertices, which duplicates the leading-order gluon-scattering
contribution \Ref{kappalo}.
To obtain the correct contribution to the coefficient $\tilde{C}^{\QCD}$
defined below \Ref{defkappa}, this leading-order contribution must be
subtracted; this subtraction can be understood as part of a systematic
matching procedure like that discussed in section \ref{sec:effective}.
More precisely, one should subtract the contribution
to \Ref{theexpression} which arises from the tree vertices 
(this corresponds to keeping only the $Z_i^{\rm tree}$ part of $Z_i$
in \Ref{refstuff2}), using the bare propagators.

The $|q^0|\approx q_\perp$ region presents some subtleties,
that are discussed in greater detail in the next section: due to the
square root singularities that appear in the vertex functions
(in the terms proportional to $X_i$, in \Ref{refstuff1}-\Ref{refstuff2}),
which enter squared in \Ref{theexpression}, the frequency integral
is potentially logarithmically divergent in the limit $q^0\to q_\perp$.
In the next section we verify that
the divergences cancel out between the lower and upper limits,
$|q^0|\to q_\perp^-$ and $|q^0|\to q_\perp^+$, although not individually.
As a consequence the integral must be evaluated
using a Cauchy principal value prescription near $q^0=q_\perp$.
Actually in the next section we show that in addition to this
Cauchy principal value integral there is
an additional $\delta(q^0-q_\perp)$ type of contribution \Ref{leftoverqperp},
giving $C_{(A),q^0=q_\perp}\simeq 0.023333$.

In addition to this, the zero-frequency leftover
\Ref{bizarreremainder} must also be included:
\bea C_{(A),q^0=0} &=&
3\pi \int_p \frac{p}{(1+p^2)^2}\left[ \frac{\tan^{-1}(p)}{p} - \frac{1}{1+p^2}\right]
\nl &=& \frac{3\pi}{32} \simeq 0.294524  \label{leftover2}
\eea

The evaluation of \Ref{theexpression} was performed
by numerical integration independently by the two authors.
The integrals giving the cut-cut, pole-cut and pole-pole
contributions are respectively four, three, and two-dimensional.
The independent evaluations used different reparametrizations of the
integration variables.
For instance, the cut-cut integration can be parameterized in terms
of the magnitudes of $\bq$ and $\br$, the angle
between them, plus one frequency variable, or in terms of $p$, $p\cdot
q$, $q_\perp^2$, and one frequency variable.
Both implementations used the Cauchy principal value prescription
near $q^0=q_\perp$ by ``folding'' the integrals in order that
the two individually divergent parts can be added together
under the integration sign and a convergent integral be
obtained.
We found satisfactory convergence in all cases,
and obtain $C_{(A),{\rm main}}\simeq 0.5918$. Combining
with Eqs. (\ref{leftover2}) and (\ref{leftoverqperp}), we thus
find $C_{(A),{\rm Im}}\simeq 0.9097$.

There exist several ways to decompose
\Ref{theexpression} into different contributions.
One way, although probably not gauge-fixing independent,
is to separate the contributions according to whether they
have tree-level or HTL interaction vertices.
Doing so, we find that the contributions involving
two HTL vertices are all relatively small, and add up to a relatively
modest $\simeq +0.14$.
There are two large contributions involving two tree interaction
vertices,
both of which come from the transverse-transverse loop:
one is the pole-pole contribution $\simeq -0.52$,
which describes the influence of the plasmon dispersion relations
on the scatterings, and the other one is the pole-cut contribution
$\simeq +0.56$, which describes scattering processes with the radiation
or absorption of a soft plasmon.
These two contributions happen to nearly cancel against each other,
so the net contribution from diagrams with two tree interaction
vertices is also relatively modest, $\simeq 0.12$.
The remainder of $C_{(A),{\rm main}}$ comes from the HTL-tree
diagrams, which add up to $\simeq +0.34$, but originate from
a large number of terms with different signs.
A large contribution comes from the
transverse-transverse cut-cut region, giving $\simeq +0.50$,
but this is largely cancelled by the transverse-transverse pole-cut
region, giving $\simeq -0.33$.  Similar cancellations happen between
the transverse-longitudinal cut-cut and pole-cut regions,
which respectively give $\simeq -0.20$ and $\simeq +0.18$.
The remainder of the HTL-tree contributions comes from the
transverse-transverse pole-pole contribution $\simeq +0.30$.
We note that the total transverse-transverse pole-pole contribution,
which we expect to be gauge invariant on its own (because this
diagram is the only place where two soft transverse plasmons can appear)
gives about $\simeq -0.20$.

\subsubsection{A subtlelty near $q^0=q_\perp$}
\label{sec:subtleties}

We now investigate in more detail the region $|q^0|\simeq q_\perp$
of \Ref{theexpression}. The purpose of this section is to verify
that the logarithmic divergences in this expression cancel out
between the lower and upper limits, $|q^0|\to q_\perp^-$ and
$|q^0|\to q_\perp^+$, so that this expression makes sense as a
Cauchy principal value integral. However, we will show that to take
such a prescription is not exactly the correct thing to do,
but that in addition there is the contribution \Ref{leftoverqperp}.

One procedure for regulating the divergences near
$|q^0|=q_\perp$ is to explicitly keep the $i\epsilon$ terms finite
in the denominators of the HTL vertex functions $M$ and $K$ \Ref{defM}:
in the time domain this regulation procedure is analogous
to placing an upper bound on the time separation between the external
legs of the self-energy diagram.
At finite time separation
no divergence is found, so that
the cancellation we find in this section means
that no significant contribution to diagram (A) arises
when the time separation between the external legs becomes large
(relative to $1/gT$.)
Our writing of \Ref{theexpression} is entirely compatible with this
regularization, since this expression follows from 
Eqs. (\ref{cutabc}) and (\ref{cutivII})
by simply decomposing each term into its real and an imaginary part;
the Wightman self-energy $\Pi^>$ is always purely real, even when
this regulator is used.

All that gets modified at finite $i\epsilon$ are the
the explicit expressions for the HTL vertex functions
given in \Ref{refstuff1}-(\ref{refstuff2}).
The only terms we need to keep track
of are those involving the function $M^{00}(Q,R)$
and its discontinuities, e.g. the terms multiplying $X_i$,
since all other terms (and propagators) are well behaved in
the kinematic region $|q^0|=q_\perp$.
Keeping the $i\epsilon$'s finite in the formulae of Appendix
\ref{app:fcts}, explicit expressions for the singular part
of the various combinations of $M^{00}(Q,R)$ and its discontinuities
that enter \Ref{theexpression} can be obtained:
\bea
K_Q &\sim& \frac{-\pi}{2p}\left[
\frac{\theta(-p\cdot q)}{\sqrt{q_\perp^2-(q^0+i\epsilon)^2}}
+\frac{\theta(p\cdot q)}{\sqrt{q_\perp^2-(q^0-i\epsilon)^2}}
\right], \nl
K_R &\sim& \frac{-\pi}{2p}\left[
\frac{\theta(-p\cdot r)}{\sqrt{q_\perp^2-(q^0-i\epsilon)^2}}
+\frac{\theta(p\cdot r)}{\sqrt{q_\perp^2-(q^0+i\epsilon)^2}}
\right], \nl
\Re K &\sim& \frac{-\pi}{4p} \left[
\frac{1}{\sqrt{q_\perp^2-(q^0+i\epsilon)^2}}
+\frac{1}{\sqrt{q_\perp^2-(q^0-i\epsilon)^2}}
\right], \nl
\Re M -\Re K &\sim&
\left[2\theta(-p\cdot q)+2\theta(-p\cdot r)-1 \right] \Re K\,.
\label{lotsofstuff}
\eea
These expressions are valid at both positive and
negative $q^0$, where the value they take is the complex conjugate
of their value at positive $q^0$.
Let us first have a look on the bracket from the first line
of \Ref{theexpression}, which multiplies $\rho(Q)\rho(R)$.
Using the fact that (because $p\cdot r=p^2-p\cdot q$) it is impossible
for both $p\cdot q$ and $p\cdot r$ to be simultaneously negative,
one can see from \Ref{lotsofstuff} that the singular behavior
of $(\Re M-\Re K)^2$ is always precisely equal to that of $(\Re K)^2$.
Thus the singular part of this bracket can be written:
\be 2(\Re K_Q)^2 - (\Im K_Q)^2 -(\Im K_R)^2 = \Re (K_Q^2+K_R^2)
\ee
a result which is explicitly well-behaved and given by a Cauchy principal
value integration%
\footnote{This follows from the standard
formula $\frac{1}{x-i\epsilon}={\cal P}\frac{1}{x} +i\pi\delta(x)$,
which is applicable here since everything it multiplies is smooth
as $q^0\to q_\perp$. \label{footnotepole}
} near $|q^0|=q_\perp$.
The bracket on the fourth line, multiplying $\Im G(Q) \Im G(R)$,
similarly yields a finite Cauchy principal value integral.
The cancellation of the divergences between the lower and upper limits
can be seen from the explicit expressions for the divergent part
of the bracket:
\bea
(\Re M - \Re K)\Re K &\sim& \frac{\pi^2}{4p^2}\frac{2\theta(-p\cdot
  q)+2\theta(-p\cdot r) -1}{q_\perp^2-q_0^2}\theta(q_\perp-q^0) \,, \nl
-\Im K_Q \Im K_R &\sim& \frac{\pi^2}{4p^2} \frac{\sgn(p\cdot q)\sgn(p\cdot
  r)}
{q_0^2-q_\perp^2} \theta(q^0-q_\perp)\,.
\eea
The two lines are opposite of each other for
all values of the momenta, as follows from the fact that
$p\cdot q$ and $p\cdot r$ are never negative at the same time.

The brackets on the second and third lines of \Ref{theexpression}
are manifestly finite when $\epsilon=0$, since the real parts of $M$ and
$K$ are only divergent for $|q^0|\to q_\perp^-$, and the imaginary parts
of the $K$'s are only divergent for $|q^0|\to q_\perp^+$:
products of these terms contain no divergence.
However, at finite $\epsilon$ the supports of the divergent
parts of these terms overlap with each other, on a region of size
$\OO(\epsilon)$. The contribution from this region remains finite
in the $\epsilon\to 0$ limit, giving rise to a
$\delta(q^0-q_\perp)$-type of contribution.
It can be extracted by just taking the
imaginary part of products of expressions from
\Ref{lotsofstuff},
using the formula mentioned in footnote \ref{footnotepole};
upon rescaling variables by $\mD$ one obtains the dimensionless
contribution to $C$ of \Ref{defkappa}:
\bea
C_{(A),q^0=q_\perp} &=& 3\pi^4 \int_p \frac{p^2}{(1+p^2)^2}
\int_Q \frac{\delta(q^0-q_\perp)}{p^2 q_\perp}\sum_i P_i\,X_i^2 
\nl && \times 
\left\{\begin{array}{l}
(\theta(-p\cdot q)-\theta(-p\cdot r)) \\ \times
\left[\Im G^{Q_i}(Q)\rho^{R_i}(R)+\rho^{Q_i}(Q)\Im G^{R_i}(R)\right]
\\
+\theta(p\cdot q)\theta(p\cdot r) \\ \times \left[
\Im G^{Q_i}(Q)\rho^{R_i}(R)-\rho^{Q_i}(Q)\Im G^{R_i}(R)\right]
\end{array}\right. \label{leftoverqperp} \nl
&\simeq& 0.0233326\,,
\eea
which turns out to be a small contribution,
most of which arising from the transverse-transverse contribution
(this result was obtained by numerical quadrature.)


\subsection{The diagram (B)}

The diagram (B) represents the expectation value of the correlator
\Ref{kappacorrel}, in which the heavy quark's Wilson line is
expanded to linear order in the $A^0$ field,
\bea
\kappa_{(B)} &=& \frac{C_H g^3}{3d_A}f^{abc}
\int_{-\infty}^\infty dt \int_0^t dt' \la
\partial^iA^{0\,a}(t) \,A^{0\,b}(t') \,\partial^i A^{0\,c}(0) \ra, \\
&=&
\frac{C_H g^3}{3d_A} f^{abc} \int_{Q,R} \frac{2i p\cdot q}{r^0+i\epsilon}
\la A^{0\,a}(Q)\,A^{0\,b}(R) A^{0\,c}(X=0) \ra \,2\pi\delta(p^0)\,,
\label{usefulB}
\eea
all fields being Keldysh $r$ fields.
The assignment of the momenta in this section is
illustrated in Fig. \ref{fig:diagrams}; $P\equiv Q+R$.
The fact that one of the fields carries zero frequency is probably
best visualized if time translation symmetry is used to move the
middle field's time argument to $t'=0$: restricting the
time argument of the leftmost field to $t>0$ (thus picking up a factor
of two), the rightmost field's time argument $t''$ in the first line of
\Ref{usefulB} is then restricted
to the range $-\infty< t'' \leq 0$. 
However, due to the antisymmetry of the group theory factor,
this range can be extended to cover the whole real axis, the
contribution from $0\leq t''<\infty$ giving zero.

\begin{FIGURE} {
\putbox{0.20}{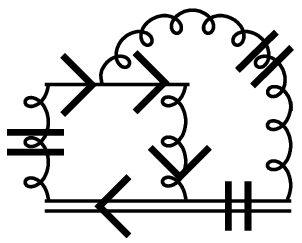}
\hfill
\putbox{0.20}{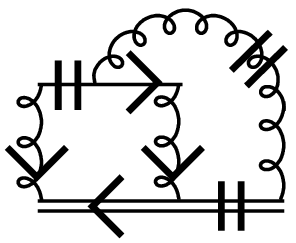}
\hfill
\putbox{0.20}{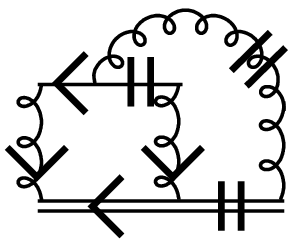}
\hfill
\putbox{0.20}{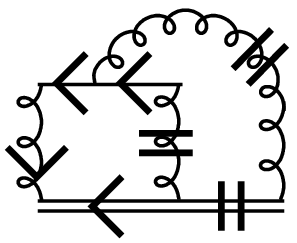}
\label{fig:diagB1}
\caption{Zoom on the $ra$ structure of the propagators in the
HTL diagrams with topology (B), when the zero frequency
gauge boson propagator (rightmost one) is cut ($rr$).
These diagrams form a telescopic sum.}
} \end{FIGURE}

Since there exists no bare interaction vertex with three $0$
Lorentz indices, this correlator only receives a contribution
from the diagram with an HTL three-point vertex.
The dominant diagrams that are
allowed by the rules of the $ra$ formalism either involve two cut ($rr$)
gluon propagators and one retarded propagator, with an $arr$ HTL vertex,
or one cut propagator and two retarded propagators, and an $aar$ HTL
vertex.  We find it convenient to organize the resulting diagrams into
two classes, according to whether the gluon propagator which carries
zero frequency is cut or retarded.
When it is cut,
one has the HTL diagrams of Fig.~\ref{fig:diagB1}, which form a
telescopic sum evaluating to:
\bea
\kappa_{(B),{\rm Re}} &=&
\frac{C_H m_D^2g^4N_c}{3} \int_{p,Q}
\frac{2i p\cdot q}{r^0+i\epsilon}
G_{rr}^{00}(p)
\int_v \left\{ \begin{array}{l} \D
-G_A^{00}(Q) \frac{1}{v\cdot Q^+}\frac{1}{v\cdot R^-} G_R^{00}(R) \\ \D
+G_R^{00}(Q) \frac{1}{v\cdot Q^-}\frac{1}{v\cdot R^+} G_A^{00}(R).
\end{array} \right.
\nl &=&  -\frac{C_H m_D^2 g^4 N_c}{3} \int_{p,q}
G_{rr}^{00}(p)
\frac{2p\cdot q}{(q^2+m_D^2)(r^2+m_D^2)}\int_v \frac{1}{v\cdot
  q^-v\cdot r^+} \label{diagB1}\,.
\eea
The second line follows from the first by means of contour integration in
the complex $q^0$ plane. The fact that one ends up with an integral
involving only zero-frequency propagators is analogous to what happened
for the real part of the gluon self-energy diagrams in section \ref{sec:re},
which we evaluated in terms of the zero Matsubara modes.
This is why we called this contribution ``$\kappa_{(B),{\rm Re}}$''.
The $v$ integration gives the function $-M^{00}(q,-r)$, which is evaluated in
\Ref{evalM}. Because it is symmetrical under $q\leftrightarrow r$
the factor $2p\cdot q$ can be traded for a $p^2$.
Upon rescaling variables by $m_D$ and scaling out the prefactor
$C_Hg^4T^2m_D/18\pi$ from \Ref{defkappa}, one obtains the dimensionless
contribution:
\bea C_{(B),{\rm Re}} &=& -6\pi^2 \int_{p,q}
\frac{p}{(1+p^2)^2(1+r^2)(1+q^2)}\frac{\pi-\cos^{-1}\left(\frac{q\cdot r}{qr}\right)}
{\sqrt{q^2r^2-(q\cdot r)^2}}
\nl &\approx& -0.0482933\,, \label{diagB1res}
\eea
in which the branch of the inverse cosine on the first line ranges from $0$ to $\pi$.
The final result was obtained by means of numerical quadrature.

\begin{FIGURE} {
\putbox{0.20}{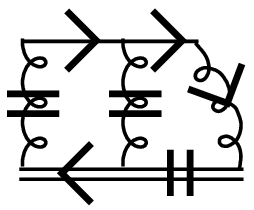}
\hfill
\putbox{0.20}{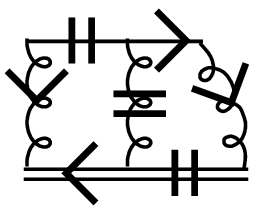}
\hfill
\putbox{0.20}{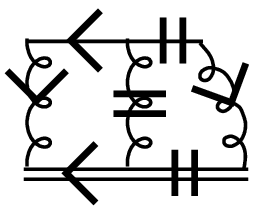}
\label{fig:diagB2}
\caption{Zoom on the $ra$ structure of the propagators in the
HTL diagrams with topology (B), when the zero frequency
gauge boson propagator (rightmost one) is retarded.  Some crossed
diagrams are not shown.}
} \end{FIGURE}

When the zero-frequency gluon propagator in \Ref{usefulB} 
is a retarded propagator, one has the HTL diagrams of
Fig.~\ref{fig:diagB2}, the evaluation of which yields:
\bea \hspace{-0.5cm} \kappa_{(B),\rm Im} = \frac{C_H m_D^2 g^4N_c}{3} \int_{p,Q}
G_A^{00}(p) \left[ \frac{i q \cdot p}{r^0+i\epsilon} - \frac{i r\cdot
    p}{q^0+i\epsilon}\right]
\times \left\{ \begin{array}{l} \D
G_{rr}^{00}(Q) G_{rr}^{00}(R) q^0 M^{00}(Q,R)^*   \\
- TG_R^{00}(Q) G_{rr}^{00}(R) K(Q,P)^* \\
+ TG_{rr}^{00}(Q) G_R^{00}(R) K(R,P)^* \\
+2TG_R^{00}(Q) G_{rr}^{00}(R) K(P,Q)\,,
\end{array} \right. \label{diagB2}
\eea
where we have symmetrized the factors from \Ref{usefulB}
under $(Q\leftrightarrow R)$.
This expression bears much similarity
to those encountered in section \ref{sec:ima} for the imaginary part of the
self-energy diagram (A).  From now on the
discussion closely parallels that given there, so we will be brief.
Indeed, the first three lines
of the brace in \Ref{diagB2} are very similar to the expressions
in \Ref{cutabc}
pertaining to cuts (i) and (ii), except that now there is only one HTL vertex
which gets cut. These expressions are also nearly in a form
suitable for numerical integration, since they are
manifestly supported on the spectral weights of the gluon propagators
(they decompose into pole-pole, pole-cut and cut-cut contributions,
like \Ref{theexpression}.)
However, just like \Ref{cutivI} from section \ref{sec:ima},
the fourth line of \Ref{diagB2} poses additional
difficulty: its support extends beyond these regions.
In section \ref{sec:ima} this difficulty was dealt with
by writing $G_{rr}(R)=(G_R(R)-G_A(R))/(r^0-i\epsilon)$,
then dropping the $G_A(R)$ term using its complex analyticity in the
upper-half $q^0$ plane, and finally using $(Q\leftrightarrow R)$ symmetry
to convert $2K(P,Q)$ into the better-behaved combination
$(K(P,Q)+K(P,R))=2\Re K(Q,R)$.
When one repeats these manipulations here, one first needs to switch the
$i\epsilon$ prescription in the $iq\cdot p/(r^0+i\epsilon)$ prefactor in
\Ref{diagB2}, in order to make it coherent with the one we wish to introduce.
This gives rise to a contribution proportional to $\delta(q^0)$;
otherwise the results are very similar.

We note that \Ref{diagB2} is manifestly real, the
contribution from $q^0<0$ being the complex conjugate of the
contribution from $q^0>0$.
Taking the real and imaginary parts of each term, as we did in
section \ref{sec:cuta},
the contribution to dimensionless $C$ \Ref{defkappa} can be written:
\bea C_{(B),{\rm Im}} &=& 6\pi \int_{p,q} \frac{p^2}{1+p^2}
\left\{ \begin{array}{l} \displaystyle
\int \frac{dq^0}{2\pi}
  \frac{1}{q_0^2}\times \left\{
\begin{array}{l} 
\rho(Q)\rho(-R) \left[ \Re M(Q,R) - \Re K(Q,P) \right]\\
+2\rho(Q)\Im G(R) \Im K(R,P) \\
+2\rho(R)\Im G(Q) \Im K(Q,P) \\
+4\left[\Im G(Q)\Im G(R) - |_{q^0=0} \right] \Re K(Q,P)
\end{array}
\right. \\ \displaystyle
+ \frac{\pi}{2pq_\perp}\left[
  G_{rr}(q)\Im G(r) + \Im G(q) G_{rr}(r)\right]/T\,.
\end{array} \right. \label{diagB2final}
\\ &\simeq& \approx -0.07338  \label{diagB2res}
\eea
The fourth line in this expression arises solely from the fourth line
of \Ref{diagB2} and the subtraction ``$|_{q^0}$'' in it means
to subtract $\Im G(Q)\Im G(R)$ evaluated at $q^0=0$;
this subtraction is convenient because it makes the integrand
well-behaved\footnote{Without this subtraction, the $i\epsilon$ prescription
in $1/(q^0+i\epsilon)^2$ would have had to be explicitly kept.}
near $q^0=0$, and it is justified by the fact that
$\int_{-\infty}^\infty dq^0 K(P,Q)/(q^0+i\epsilon)^2=0$.
A part of the ``leftover'' from $q^0=0$ on the last line of \Ref{diagB2final}
arises from the manipulations described in the preceding paragraph, concerning
the fourth line of \Ref{diagB2},
and another part of it arises when the real part of the square bracket
in \Ref{diagB2} is taken, since this contains a $\delta(q^0)$ term. 

Explicit expressions for the
functions $\Re M$, $\Re K$ and $\Im K$ that appear
in \Ref{diagB2final} can be obtained by setting $X=1$ and $Y_Q=Y_R=Z=0$
in \Ref{refstuff1} and \Ref{refstuff2} of section \ref{sec:finala}.
We evaluated \Ref{diagB2final} by numerical quadratures;
the integration splits into pole-pole, pole-cut and cut-cut contributions
(plus the zero-frequency leftover) and their evaluation
is very similar to the self-energy diagram (A).
There is one important subtlety, though:
neither the pole-cut, cut-cut nor $q^0=0$ contributions
are individually well-defined, as they are all (logarithmically)
ultraviolet divergent in the limit of large $r$, $q$ fixed
(taking $r>q$, for definiteness). However their sum is UV convergent;
here we skip the explicit verification of this fact.
What this implies is that these contributions must be added under the
integration sign, e.g. the content of the large brace in
\Ref{diagB2final} must be added up before the integration over
spatial $q$ and $p$ is performed.
The largest part of \Ref{diagB2res} originates from the pole-pole
region: thus diagram (B) mostly describes scatterings
against longitudinal plasmons (this diagram is an interference
term between the Coulomb and Compton channels for
longitudinal plasmon scattering.)

\subsection{Self-energy (C): imaginary part}
\label{sec:imc}

The three different diagrams
contributing to the cut self-energy diagram of type (C)
are shown in Fig. \ref{fig:cutC}.
A direct evaluation of these diagrams would yield:
\bea \Pi^{>\,00}_{(b)}(P) &=& g^2N_c m_D^2T \int_Q G_{rr}^{\mu\nu}(Q)
\int_v \,v_\mu v_\nu \left[
\frac{2\pi\delta(v\cdot P)}{v\cdot P^- v\cdot (P+Q)^-}
\right. \nl && \hspace{1.5cm} \left. 
+\frac{2\pi\delta(v\cdot (P+Q))}{v\cdot P^- v\cdot P^+}
+\frac{2\pi\delta(v\cdot P)}{v\cdot P^+ v\cdot (P+Q)^+}
\right] \label{diagC1}\,.
\eea
However, this contains manifestly ill-defined factors such as
$\delta(v\cdot P)/v\cdot P^-$. These ill-defined expressions are a
typical example of pinch singularities and are due
to cuts (i) and (iii) of Fig. \ref{fig:cutC},
which are attempting
to provide self-energy corrections to an on-shell propagator.
However, as is well-known, such pinch singularities always cancel out,
and in the end one typically obtains expression possessing a
``gain-term, loss-term'' type of structure
characteristic of Boltzmann-like transport equations
(see e.g. \cite{greiner}).
In our case the simplest way to regulate the pinch singularities
is to use the space-time description of the four-point HTL
with two external Keldysh $a$ indices%
\footnote{The result we obtain with this regulator
agrees with what one would
obtain by regulating the pinch singularities by resumming self-energies
into the propagators.
Indeed, the two regularization methods only differ on frequency scales
of order the collisional width (damping rate) $\Gamma\sim g^2T$,
whilst the effects we are looking at take place on the
frequency scale $gT$.
}, which is simply given
by an adjoint Wilson line along the trajectory of the light particle \cite{htlpaper}:
\be \Pi^{>\,00}_{(b)}(P) = -g^2 N_c m_D^2 T \int_v v_\mu v_\nu
\int_{-\infty}^\infty dt\,
e^{-it v\cdot P} \int_0^t dt' \int_0^{t'} dt''
\int_Q e^{-i(t'-t'')v\cdot Q}\,G_{rr}^{\mu\nu}(Q). \label{diagC2}
\ee
Here $t$ is the time separation between the two endpoints
of the Wilson line and the integrand
gives the expectation value of a Wilson line in a fluctuating
background gauge field, the Wilson line being expanded to second
order in the background field and the fluctuations being described by
the $G_{rr}$ correlator.
Performing the integrations over the time arguments and using
the symmetry of $G_{rr}(Q)$ to drop
terms which are odd under $Q\to -Q$,
one obtains the well-defined integral:
\be \Pi^{>\,00}_{(b)}(P) = g^2 N_c m_D^2 T
\int_v v_\mu v_\nu \int_Q G_{rr}^{\mu\nu}(Q) \,2\pi
\frac{\delta(v\cdot (Q+P))-\delta(v\cdot P)}{(v\cdot Q)^2}\,.
\label{usefulC}
\ee
which has a transparent structure as the sum
of a gain term and a loss term.

\begin{FIGURE}[t] {
\putbox{0.32}{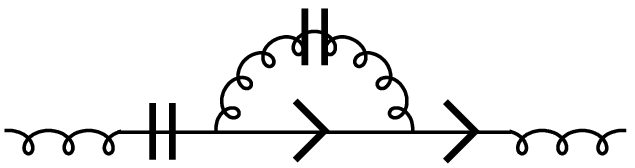}
\hfill
\putbox{0.32}{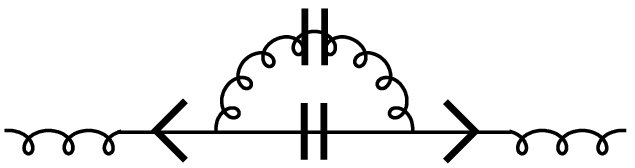}
\hfill
\putbox{0.32}{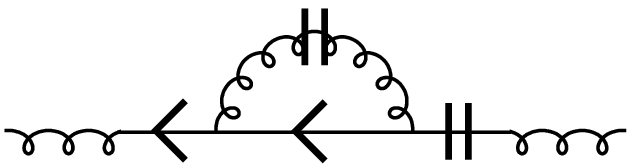}
\label{fig:cutC}
\caption{Zoom on the $ra$ structure of the propagators in the
HTL diagrams with topology (C).}
} \end{FIGURE}

To evaluate \Ref{usefulC} we first rotate
$\bq$ so that its $z$ axis lies aligned with $\bv$.
The remaining angular integration is over the dot product $u\equiv
\bv\cdot \bp/p$. Although $u$ integration naturally ranges between -1 and 1,
by noting that one would obtain zero if it were extended
to cover the whole real axis,
one can trade it for an integration over $|u|\geq 1$,
which we find more convenient for numerical purposes. Upon performing the $p$
integration \Ref{pintegration}, rescaling variables by $\mD$
and factoring out a numerical prefactor,
one obtains the contribution to dimensionless $C$ \Ref{defkappa}:
\bea C_{(C)} &=&
6\pi \int_p \frac{-1}{(1+p^2)^2} \int_1^\infty \frac{du}{u^2}
\int_q
\left[ G_{rr}^{00}(up+q_z,q) + (1-q_z^2/q^2) G_{rr}^T(up+q_z,q)
\right]/T \label{diagC4}
\nl &\simeq& -0.132916 \label{diacCres}
\eea
where the correlators $G_{rr}$ are to be evaluated in the soft
$p^0$ approximation $G_{rr}(P)=(G_R(P)-G_A(P)) T/q^0$.
Although \Ref{diagC4} can be simplified somewhat (for instance
either the $u$ or the $p$ integration, at fixed $up$, can be done by
hand), we had to evaluate it using numerical quadrature.
One encounters pole and cut types of contributions in the transverse
and longitudinal channels.
The result turns out to be almost completely determined by the
contribution from the transverse pole, which produces $-0.113353$ by itself.
Thus the important physics described by this diagram is not that of
overlapping scattering events, but rather that of
tree-level Coulomb scattering processes accompanied
with the emission or absorption of a soft transverse gluon
(however we do not expect
diagram (C) to give a gauge-invariant account of these
processes by itself,
as the emitted transverse gluon may also be emitted at the
exchange gluon, yielding diagrams that are included in (A).)

\subsection{The diagram (D)}
\label{sec:diagd}

This diagram involves two insertions of a fluctuating gauge field
along the heavy quark's adjoint Wilson line \Ref{kappacorrel},
and can be written as:
\be \kappa_{(D)} = \frac{-C_Hg^4N_c}{3} \int_{P,Q} \int_{-\infty}^\infty
dt \int_0^t dt' \int_0^{t'} dt'' \,e^{-it p^0} e^{-i(t'-t'')q^0}
p^2\, G_{rr}^{00}(P) G_{rr}^{00}(Q).
\ee
The structure of this expression is very similar
to that of \Ref{diagC2}, met in considering diagram (C),
the only difference now being
that the Wilson lines lie along the static trajectory of the heavy
quark, as opposed to the light-like trajectories of the light plasma
particles. Upon performing the time integrations, rescaling variables
by $\mD$ and scaling out
the numerical prefactor $C_Hg^4T^2m_D/18\pi$, we obtain the dimensionless
contribution:
\bea C_{(D)}&=& 6\pi \int_{p,q} \int \frac{dq^0}{2\pi} p^2
\frac{G_{rr}^{00}(q^0,p)-G_{rr}^{00}(0,p)}{q_0^2}
G_{rr}^{00}(q^0,q) \label{usefulD} \nl
&\simeq & 0.0675263 \,.
\eea
This was obtained by means of numerical quadrature; the integral splits into
pole-pole ($\simeq 0.0474$), pole-cut ($\simeq 0.0261$) 
and cut-cut ($\simeq 0.0097$) contributions, which respectively
require one-, two-, and three-dimensional integrals.
In addition there is a contribution coming from the zero-frequency
term $G_{rr}^{00}(0,p)$ ($\simeq -0.0156$), whose support
in the $q^0>p$ region needs to be handled separately; these integrals
posed no particular difficulty.

This exhausts the diagrams contributing the the momentum diffusion
coefficient at next-to-leading order in QCD.
Taking the sum of the numbers we obtained we find $\tilde{C}\simeq 1.4946$,
with $\tilde{C}$ as in \Ref{defkappa}.

\section{\Nfour\ super Yang-Mills}
\label{sec:SYM}

Our setup for \Nfour\ SYM was described in section \ref{sec:resSYM}.
To perform the next-to-leading order calculation we need the appropriate
generalization of the force-force correlator and Wilson line appearing
in \Ref{kappacorrel}:
\be
\kappa^{\SYM} = \frac{\lambda^2}{6d_A}\int_{-\infty}^\infty dt\,
\la \partial^i (A^0+\phi)^a(t)
\left[{\cal P} e^{\int_0^t dt' [A^0+\phi,\,\cdot]}\right]^{ab} \partial^i (A^0+\phi)^b(0)
\ra \label{correlSYM}\,,
\ee
where $\phi$ stands for the (canonically normalized) real scalar field
of SYM which couples to the nonrelativistic heavy quark.
The derivation of this formula is identical
to that of \Ref{kappacorrel}; one must account for the influence of the
scalar field
on the eikonal propagation of the heavy quark and on the force it feels%
\footnote{Had we written the SYM version of \Ref{correl1} instead of
that of \Ref{kappacorrel},
the force term would involve the manifestly gauge-covariant combination
$(F^{i0}+D^i\phi)$.
The spatial gauge fields $A^i$ enter this expression in just such a way
as to give total time derivatives which can be dropped,
as discussed above \Ref{kappacorrel}.}.
In expanding the correlator \Ref{correlSYM} into powers of $A^0$ and $\phi$
only the terms with even powers of $\phi$ have to be kept, since correlators
of odd powers of $\phi$'s vanish by virtue of the global SU(4)
R-symmetry of \Nfour\ SYM.
The power-counting is the same as for QCD:
we need to retain the diagrams with two soft loops.
The nonvanishing ones, which we did not previously
compute in dealing with QCD, are depicted in Fig.~\ref{fig:SYMNLO}.

Interestingly, these diagrams are significantly
easier to evaluate than those of the preceding section,
due to the simplicity of HTL amplitudes involving soft scalars:
the HTL scalar self-energy reduces to a  \emph{momentum-independent}
mass shift $m_S^2=\lambda T^2$, and there 
exists no HTL effective vertex with external scalars \cite{braatengen}.
These features are generic to theories containing scalar fields
(with no cubic scalar self-interactions).
Physically, the absence of an imaginary part (Landau cut) in the soft
scalar two-point function
(from which the momentum-independence of the real part of
the self-energy follows, by a Kramers-K\"onig relation)
can be attributed to the spin-suppression of the
processes by which soft virtual scalars would be produced
by the small-angle scattering of light plasma particles with spin.
Indeed these processes, namely Yukawa scattering of fermions
and gluon-scalar conversion, both involve a
change in the helicity of the light particle.
Since the soft virtual scalar carries no polarization tensor,
the matrix elements for these processes must
be proportional to the deflection angle, which is $\OO(g)$ in the HTL limit.
Exactly the same mechanism prevents a background scalar field
from interfering with the eikonal propagation of relativistic
particles with spin, thus suppressing the insertion of external scalar
legs onto existing HTL amplitudes.

\begin{FIGURE} {
    \includegraphics[width=13cm,height=7.0cm]{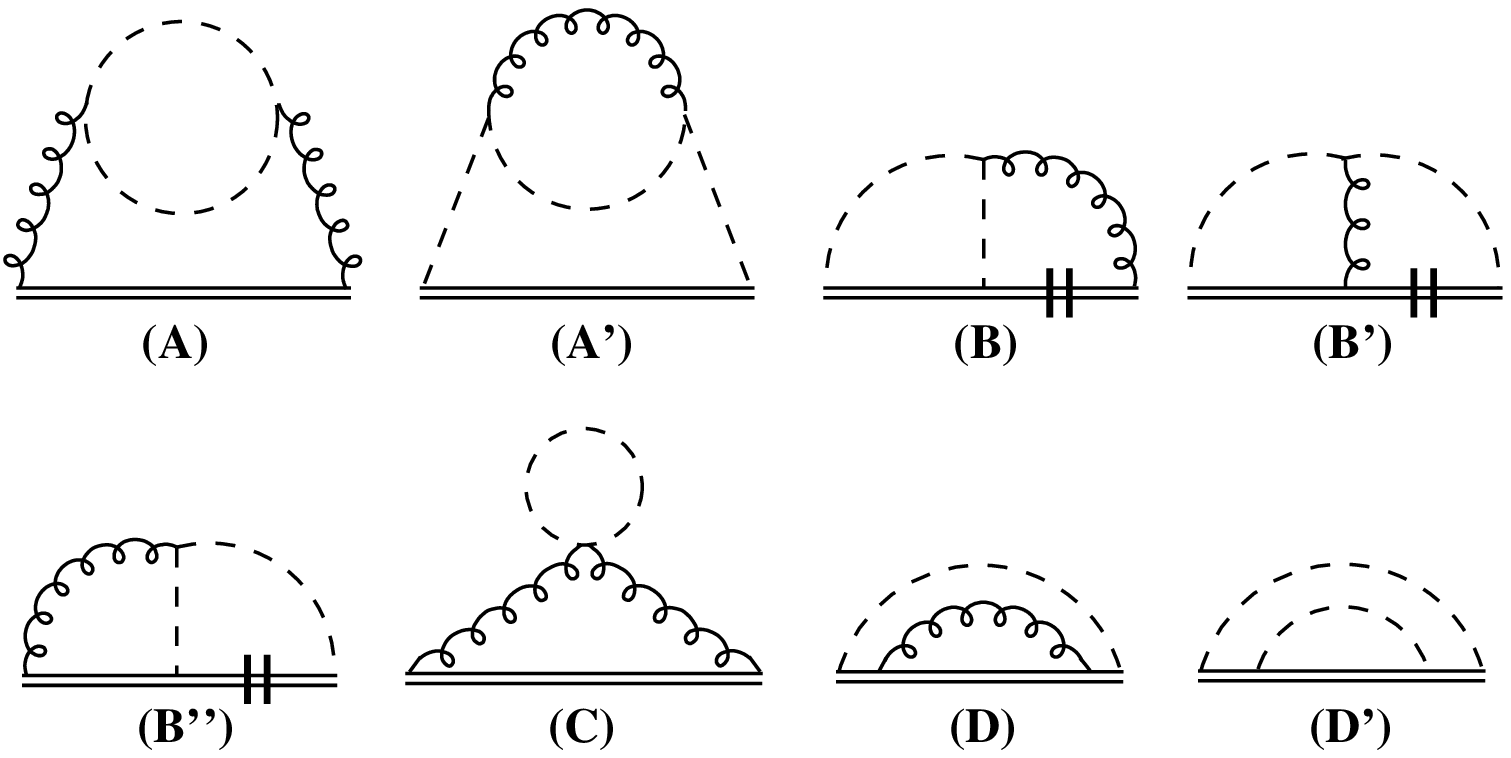}
\label{fig:SYMNLO}
\caption{Additional diagrams that contribute to the next-to-leading
  order momentum diffusion coefficient in \Nfour\ SYM.
Not shown, one permutation of (D).}
} \end{FIGURE}

Perhaps among the most conceptually transparent sources of NLO corrections from
soft scalars is that stemming from the imaginary part of the gluon
self-energy diagram (A) of Fig.~\ref{fig:SYMNLO}.
This diagram produces a negative correction, due to the
reduced phase space available for scattering against a massive scalar.
Actually, this diagram naturally combines with diagram (D') and
with the part of diagram (B)
in which the zero-frequency gluon propagator
(which is the rightmost propagator) is a retarded propagator:
their sum forms a single logical unit, which accounts for
the effect of the scalar mass to the tree processes described
by the second line of \Ref{SYMLO}:
\bea \dksym&=&\frac{\lambda^2}{24\pi^3} \int_0^\infty q^3dq
\int_{q/2}^{\infty} \frac{k^2 dk}{dE_k/dk} \,n_B(E_k)(1+n_B(E_k)) 
\nl && \hspace{2cm} \times \left. 
\left[\frac{5}{(q^2+\mD^2)^2} +
\left(\frac{1}{q^2+\mD^2}-\frac{1}{2E_k^2}\right)^2\right]
\right|_{E_k=k}^{E_k=\sqrt{k^2+\mS^2}} \,.
\eea
Here it is most convenient to trade the $k$ integration for an
integration over $E_k$:
the integrand then becomes independent on the functional form of $E_k$,
and the massive and massless results only differ due to the
different integration bounds at small $E_k$.
The NLO correction arises from the region $q\sim gT$, in which case
the $E_k$ integration
ranging from $(q/2)$ to $\sqrt{(q/2)^2+\mS^2}$ can be
done within the approximation $n_B(E_k)(1+n_B(E_k))=T^2/E_k^2$.
Upon rescaling variables by $\mD$ and factoring out
$\lambda^2T^2\mD/36\pi$, one obtains the following dimensionless contribution:
\bea \dcsym &=& -\frac{3}{2\pi^2} \int_0^\infty q^3dq
\left[ 3 \frac{\sqrt{q^2+2}-q}{(q^2+1)^2} -
  2\frac{\sqrt{q^2+2}-q}{q\sqrt{q^2+2}(q^2+1)}
+\frac{2}{3} \left(\frac{1}{q^3}-\frac{1}{(q^2+2)^{3/2}}\right)\right]
\nl &=& -\frac{3}{2\pi^2} \left( \frac{-31 \sqrt{2}}{6} + \frac{13 \pi}{4}
       - \frac{\cosh^{-1}(\sqrt{2})}{2} \right)
\simeq -0.37429   
\label{SYMI}
\eea

The real part of the zero-frequency self-energy diagram (A) gives zero at
NLO, as can be seen using the argument employed in section \ref{sec:re}:
because the interaction vertices are proportional to the loop frequency,
in the imaginary time formalism the diagram is saturated by the
non-zero Matsubara frequencies, for which no $\OO(g)$ corrections arise.
The tadpole
diagram (C) produces a (negative) momentum-independent shift to $\mD^2$:
\be \delta \mD^2= 6\lambda T\int \frac{d^3q}{(2\pi)^3} \left[
  \frac{1}{q^2+\mS^2}-\frac{1}{q^2}\right] = -\frac{3\sqrt2}{4\pi}
\lambda T \mD\,,
\ee
which induces a positive shift to $C$,
\be \dcsym = 6\pi \int \frac{d^3p}{(2\pi)^3} p^2 G_{rr}^{00}(p)
\frac{3\sqrt2}{2\pi(1+p^2)} = \frac{9\sqrt 2}{8\pi} \simeq 0.50643
\label{SYMII}
\ee

A more or less related virtual contribution arises from the terms in (B)
in which the zero-frequency gluon propagator is cut:
\be \dksym =  \frac{\lambda^2}{6} 2i\int_{p,Q} p^2G_{rr}(p)
\left[  G_{rr}^{(S)}(Q)G_R^{(S)}(R)+G_{R}^{(S)}(Q)G_{rr}^{(S)}(R)\right]\,.
\label{dummySYM}
\ee
The pattern of propagators in the bracket is the same as usually
arises in the calculation of one-loop retarded self-energies, and
because the corresponding external momentum $P$
carries zero frequency,
one finds that the dominant result at soft $p$ can be expressed in terms
of the contribution from the zero Matsubara frequency.
This can be derived from \Ref{dummySYM}
by writing $G_{rr}(Q)=(G_R(Q)-G_A(Q))T/(q^0-i\epsilon)$ in the first
term, and using analyticity in the upper-half $q^0$ plane to drop the
$G_A(Q)$ contribution, thus turning this term into
$G_R(Q)G_R(R)T/(q^0-i\epsilon)$. Similar manipulations on the second
term of the bracket yield $G_R(Q)G_R(R)T/(r^0-i\epsilon)$, which cancels
against the first leaving only a $\delta$-function at $q^0=0$.
Rescaling variables by $\mD$ one thus obtains the dimensionless contribution:
\be
\dcsym = 12\pi^2 \int_{p,q} \frac{p}{(p^2+1)^2}
\frac{1}{(q^2+\frac12)(r^2+\frac12)} 
= \frac{3}{2\pi} \int_0^\infty \frac{p^2 \,dp}{(1+p^2)^2}
\tan^{-1} \frac{p}{\sqrt{2}}
\simeq 0.37044 \,.
\label{SYMIII}
\ee

Corrections to the real part of the scalar self-energy  
are irrelevant at NLO, due to the vanishing of the imaginary part of the
HTL scalar self-energy.
The corrections (A') to the imaginary part of the scalar self-energy
naturally combine
with diagrams (B'), (B'') and (D) to form a single logical unit
(in diagrams (B') and (B'') the zero-frequency scalar propagator must be
a retarded
propagator, again because the imaginary part of the HTL
scalar self-energy vanishes.)
Together they describe the effects of the scalar mass and
plasmon dispersion relation on tree-level gluon-scalar scattering,
as well as new processes, involving physical longitudinal plasmons
in the external states or overlapping scattering events (associated with
gluon propagators in the Landau cut.)
Processes involving longitudinal gluons also occur in a Compton-like
channel, which interferes with the scalar-exchange channel: this is what
diagrams (B'), (B'') and (D) account for.
Using formulae \Ref{usefulB} and \Ref{usefulD}
these diagrams add up to:
\bea
\dcsym\! &=&\! 6\pi\!\! \int_{p,Q} \!\! G_{rr}^{(S)}(R)\!\left[
4p^2G_{rr}^T(Q)\frac{p^2-\frac{(p\cdot q)^2}{q^2}}{(p^2+\frac12)^2}
+p^2G_{rr}^{00}(Q) \left(\frac{q^0}{p^2+\frac12}-\frac{1}{q^0}\right)^2
-\frac{q^2}{q_0^2}G_{rr}^{00}(q,0)
\right] \nl &\simeq& -0.31086 \,.
%
%
\label{SYMIV}
\eea
The contribution from the transverse
gauge field is linearly divergent at large momenta,
where it degenerates to the tree-level contribution which was
already included in \Ref{SYMLO}; this must be subtracted.
What we have integrated numerically is the
difference between \Ref{SYMIV} and the same expression
evaluated with the tree $G_{rr}$ propagators,
which is given by a convergent integral.
The integral splits into pole-pole and pole-cut contributions,
with the largest contribution arising from
the transverse gluon pole, which gives $\simeq -0.283$.

This exhausts the list of NLO contributions whose origin is proper to
the SYM theory.
Taking the sum of Eqs.~\ref{SYMI}, \ref{SYMII}, \ref{SYMIII} and
\ref{SYMIV} we obtain the next-to-leading order coefficient
$\tilde{C}^{\SYM} = 0.19172$.

\centerline{\bf Acknowledgements}

\noindent
This work was supported in part by
the Natural Sciences and Engineering Research Council of Canada.

\begin{appendix}

\section{The functions $M^{00}(Q,R)$ and $L(Q)$}
\label{app:fcts}

The functions $M^{00}(Q,R)$ and $L(Q)$ have appeared in some form or the other
in previous work (\cite{braatendamping} \cite{braatenyuan} \cite{FrenkelTaylor}, to cite a few.)
The function $L(Q)$ is familiar from the longitudinal HTL gluon self-energy:
\bea L(Q)&=& \int \dv \frac{1}{v\cdot Q^-} \nl
&=& \frac{-1}{2q} \ln\left( \frac{q^0 + q+i\epsilon}{q^0-q+i\epsilon} \right).
\eea
The branch of the logarithm is such that this function is real
for time-like $Q$, and acquires a positive imaginary part at
space-like $Q$, the so-called Landau cut.

To understand the analytic structure of $M^{00}(Q,R)$ we find convenient
to express it as a sum of two terms,
\be M^{00}(Q,R) = M^{00}(P,Q) + M^{00}(P,R)\,, \label{decomposed}
\ee
and to analyze those two terms separately.
Actually, the splitting of \Ref{decomposed} as a sum of two terms
is precisely how the function
$M^{\mu\nu}(Q,R)$ arose in the first place, in the HTL
effective vertices of Fig. \ref{fig:blocks}, so in a sense
it is rather natural.
In \cite{FrenkelTaylor} an expression for the (Lorentz-invariant) function
$M^{00}(P,Q)$ was obtained
by first combining the denominators $1/v\cdot P^-$ and $1/v\cdot Q^-$ by
means of standard Feynman parameterization:
\bea M^{00}(Q,P) &=& \int_v \frac{1}{v\cdot P^- v\cdot Q^-}
\label{defm00} \\ &=&
\int_0^1 dx \int_v
\frac{1}{\left[ x\,v\cdot Q + (1-x)v\cdot P -i\epsilon \right]^2}
= \int_0^1 dx \frac{-1}{ \left(xQ+(1-x)P\right)^2 -i\epsilon q^0 } \nl
&=& \left.\frac{-1}{2\sqrt{-\Delta}}
\ln\left(\frac
{Q\cdot P + \sqrt{-\Delta}}
{Q\cdot P - \sqrt{-\Delta}}
\right) \right|_{q_0\equiv q^0+i\epsilon} \label{mmajor}\,,
\eea
where:
\be \Delta= Q^2P^2-(Q\cdot P)^2.
\ee

Proper care must be given to the $i\epsilon$
prescriptions in \Ref{mmajor}, since we are interested in this function
and its discontinuities at physical values of the momenta.
The correct procedure follows from noting that, from its definition,
the function $M^{00}(Q,P)$ is analytic in the upper half complex $q^0$
plane. Its integral representation is unambiguous for $Q$ time-like,
in which case it has no discontinuity across the real axis,
as a function of $q^0$, with fixed $p^0$, $\bp$ and $\bq$.
Using a change of variables $\bv\to -\bv$ in the integral representation
\Ref{defm00}, one shows that flipping the sign of $q^0$ in $M^{00}(Q,P)$
is equivalent to complex conjugation. Explicit expressions, when
$q^0\geq 0$,
are:
\bea \hspace{-0.5cm} M^{00}(Q,P)&=& \left\{
\begin{array}{l@{\hspace{-1em}}l} \D
\frac{-1}{2p\sqrt{q_0^2-q_\perp^2}}\left[\ln\left(
\frac{p\cdot q + p\sqrt{q_0^2-q_\perp^2}}
{-p\cdot q + p\sqrt{q_0^2-q_\perp^2}}\right) +i\pi\right],
&\hspace{1.5em} q^0>q, \\ \D
\frac{-1}{2p\sqrt{q_0^2-q_\perp^2}}\left[\ln\left(
\frac{p\cdot q + p\sqrt{q_0^2-q_\perp^2}}
{p\cdot q - p\sqrt{q_0^2-q_\perp^2}}\right) \right.& \\
\hspace{4.5cm} \left. +2\pi i\,\theta(-p\cdot q)
\parbox[t][3ex]{1pt}{}\right],&
q_\perp < q^0 < q, \\ \D
\frac{-1}{p\sqrt{q_\perp^2-q_0^2}} \left[ \tan^{-1}\left(
\frac{p\sqrt{q_\perp^2-q_0^2}}{p\cdot q}\right)\right],&
0< q^0 < q_\perp,
\end{array}\right. \label{evalM}
\eea
where all logarithms are real,
and the arctangents range from $0$ to $\pi$.

The equations \Ref{evalM} can be understood from \Ref{mmajor} as follows.
When $q^0>q$, the magnitude of the square root
is automatically larger than $|q\cdot p|$, thus the denominator
in the logarithm in \Ref{mmajor} is negative.
Since this denominator has a small
negative imaginary part, the logarithm acquires a positive
phase $+i\pi$. This choice of branch can be
verified from the large $q^0$ limit, where $M^{00}(Q,R)\to -i\pi/2pq^0$
according to its definition \Ref{defm00}.
When $q^0$ crosses $q$ ($Q$ becomes spacelike),
the square root becomes equal to $|q\cdot p|$, hence either
the numerator or the denominator of the
logarithm vanishes. Which one vanishes depends on the sign
of $q\cdot p$, explaining the appearance of the
$\theta(-q\cdot p)$ function in the second case. Finally, when
$q^0< q_\perp$, the square root becomes imaginary and
the logarithm goes over to an arctangent.

The function $K^{00}(Q,P)$ is (half)
the discontinuity of $M^{00}(Q,P)$ across the real
$q^0$ axis, and can be extracted from the previous results
by writing $K(Q,P)=(M(Q,P)+M(-Q,P))/2$:
\bea K^{00}(Q,P)&=& \left\{ \D
\begin{array}{l@{\hspace{1cm}}l}
0, & q< q^0, \\ \D
\frac{i \pi\, \sgn(p\cdot q)}{2p\sqrt{q_0^2-q_\perp^2}},
& q_\perp < q^0 < q, \\ \D
\frac{-\pi}{2p\sqrt{q_\perp^2-q_0^2}},
& 0< q^0 < q_\perp.
\end{array}\right. \label{evalK}
\eea

\end{appendix}

\end{document}